%% file: main.tex
\documentclass[final,authoryear]{elsarticle}
\usepackage[margin=1in]{geometry}

\usepackage{amssymb}
\usepackage{amsmath}
\usepackage{amsthm}

\usepackage{bm}
\usepackage{bbm}
\usepackage{todonotes}
\usepackage{enumitem}
\usepackage{subcaption}
\usepackage{mdframed}

\newtheorem{theorem}{Theorem}
\newtheorem{proposition}[theorem]{Proposition}

\theoremstyle{definition}
\newtheorem{definition}{Definition}
\newtheorem{assumption}{Assumption}
\newtheorem{example}{Example}

\newcommand{\Nat}{{\mathbb{N}}}
\newcommand{\Real}{{\mathbb{R}}}

\newcommand{\N}{{\mathcal{N}}}
\newcommand{\M}{{\mathcal{M}}}
\newcommand{\X}{{\mathcal{X}}}
\newcommand{\R}{{\mathcal{R}}}
\newcommand{\K}{{\mathcal{K}}}

\newcommand{\U}{{\mathcal{U}}}
\newcommand{\T}{{\mathcal{T}}}

\newcommand{\I}{{\mathcal{I}}}
\newcommand{\J}{{\mathcal{J}}}
\newcommand{\DRA}{{\mathfrak{D}}}

\DeclareMathOperator*{\E}{\mathbb{E}}

\journal{Annual Reviews in Control}

\makeatletter
\def\ps@pprintTitle{%
 \let\@oddhead\@empty
 \let\@evenhead\@empty
 \let\@oddfoot\@empty
 \let\@evenfoot\@empty
}
\makeatother

\begin{document}

\begin{frontmatter}



\title{A Vision for Trustworthy, Fair, and Efficient Socio-Technical Control using Karma Economies} 


\author[ifa,idsc]{Ezzat Elokda} 
\author[idsc]{Andrea Censi}
\author[idsc]{Emilio Frazzoli}
\author[ifa]{Florian D\"orfler}
\author[ifa]{Saverio Bolognani}

\affiliation[ifa]{organization={Automatic Control Laboratory, ETH Zurich},
            country={Switzerland}}
\affiliation[idsc]{organization={Institute for Dynamic Systems and Control, ETH Zurich},
country={Switzerland}}

\begin{abstract}
Control systems will play a pivotal role in addressing societal-scale challenges as they drive the development of sustainable future smart cities. 
At the heart of these challenges is the trustworthy, fair, and efficient allocation of scarce public resources, including renewable energy, transportation, data, computation, etc..
Historical evidence suggests that monetary control -- the prototypical mechanism for managing resource scarcity -- is not always well-accepted in socio-technical resource contexts.
In this vision article, we advocate for \emph{karma economies} as an emerging non-monetary mechanism for socio-technical control.
Karma leverages the repetitive nature of many socio-technical resources to jointly attain trustworthy, fair, and efficient allocations; by budgeting resource consumption over time and letting resource users ``play against their future selves.''
To motivate karma, we review related concepts in economics through a control systems lens, and make a case for a) shifting the viewpoint of resource allocations from \emph{single-shot and static} to \emph{repeated and dynamic} games; and b) adopting \emph{long-run Nash welfare} as the formalization of ``fairness and efficiency'' in socio-technical contexts.
We show that in many dynamic resource settings, karma Nash equilibria maximize long-run Nash welfare.
Moreover, we discuss implications for a future smart city built on \emph{multi-karma economies}: by choosing whether to combine different socio-technical resources, e.g., electricity and transportation, in a single karma economy, or separate into resource-specific economies, karma provides new flexibility to design the \emph{scope of fairness and efficiency}.

\end{abstract}



\begin{keyword}
Systems and Control for Societal Impact \sep Control for Smart Cities \sep Cyber-Physical Human Systems \sep Mechanism Design \sep Social Choice Theory \sep Karma Economies



\end{keyword}

\end{frontmatter}


\input{sections/introduction}
\input{sections/pillars}
\input{sections/problem}
\input{sections/karma}
\input{sections/outlook}
\input{sections/conclusion}

\section*{Acknowledgments}
We would like to thank Heinrich H. Nax for the many fruitful discussions and pointers in the economics literature.
Research supported by NCCR Automation, a National Centre of
Competence in Research, funded by the Swiss National Science
Foundation (grant number 51NF40\_80545).


\bibliographystyle{elsarticle-harv} 
\bibliography{main}

\appendix
\input{sections/proofs}

\end{document}

%% file: sections/introduction.tex
\section{Introduction}

Since its inception, the field of control systems has been a driving force for much of the technological progress that shaped the quality of modern human life~\citep{bennett2002brief}.
From automating large-scale mass-production lines, to bringing reliable and stable energy to every household,
there is an invisible controller behind many of our basic modern comforts.
But these comforts have not been without a cost: population over-growth and unsustainable consumption of natural resources have culminated in the grand challenge of \emph{climate change} that threatens our very existence.
As a consequence, world leaders have declared a fight against climate change, including the EU which committed to an ambitious target to reduce greenhouse gas emissions by 40\% compared to 1990 by 2030, and achieve net-zero emissions by 2050~\citep{ec2019going}.

\begin{figure}[!tb]
    \centering
    \includegraphics[width=0.75\textwidth]{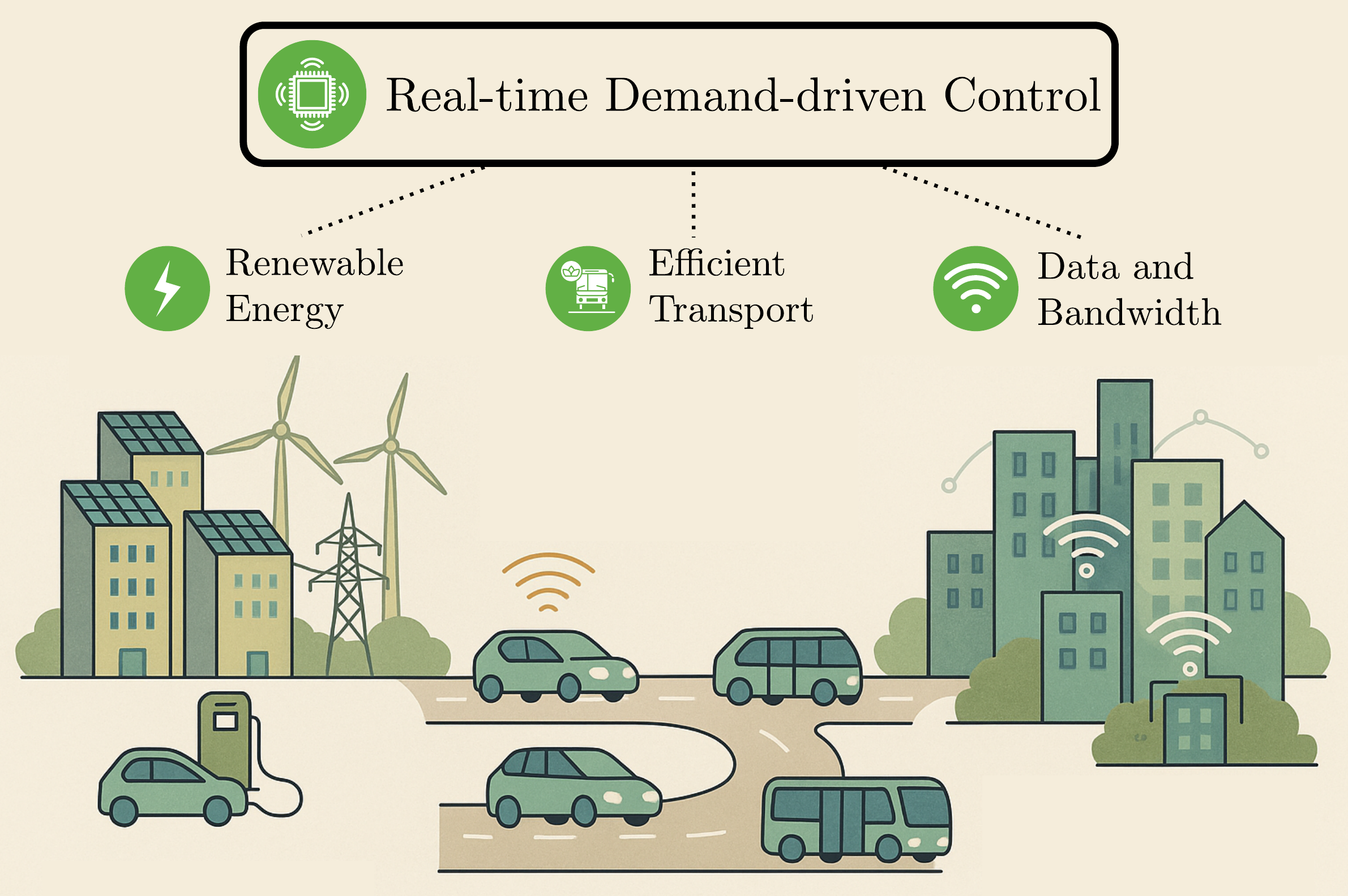}
    \caption{Vision of the future smart city: automated, efficient, and sustainable.}
    \label{fig:smart-city}
\end{figure}

\begin{figure}[!b]
    \centering
    \includegraphics[width=0.9\textwidth]{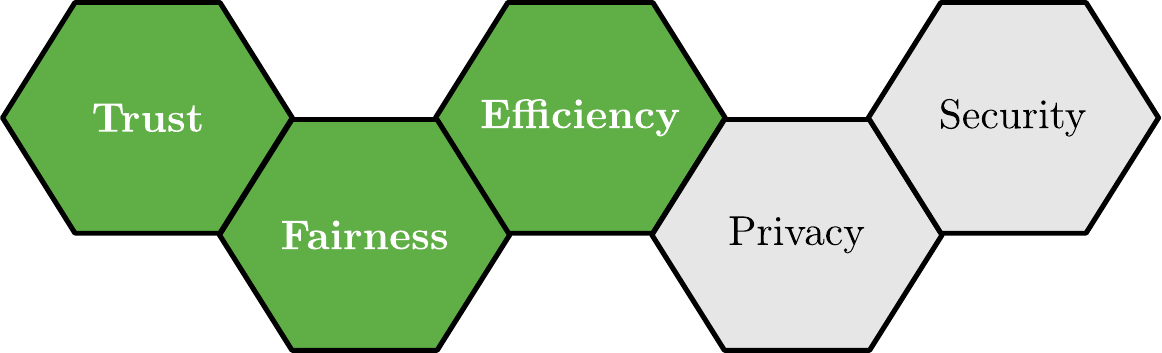}
    \caption{There are many important human factors to consider in socio-technical control design. Karma economies address the core factors of \emph{trust}, \emph{fairness}, and \emph{efficiency}.}
    \label{fig:socio-technical-factors}
\end{figure}

The control systems community has a pivotal role to play in addressing these societal-scale challenges~\citep{annaswamy2024control,khargonekar2024climate}.
Figure~\ref{fig:smart-city} shows a typical vision of the future smart city: automated, efficient, sustainable, and driven by advanced control algorithms.
At the heart of this vision lies the large-scale and automated allocation of public resources to the human citizens, whether it be electricity for household needs or for charging electric vehicles, roads and public transit, water resources, or communication bandwidth.
Also central to this vision is the promise of \emph{efficiency}: by adapting resource allocations to user demands in real-time, more critical demands can be satisfied with fewer available resources.
However, this performance-driven vision frequently overlooks several equally, if not more, important human and societal factors, a subset of which are portrayed in Figure~\ref{fig:socio-technical-factors}~\citep{annaswamy2024control}.
The control systems community bears a responsibility to rigorously and holistically consider these societal factors, among which we identify \emph{trust} and \emph{fairness} as most critical in addition to \emph{efficiency}; since they pose fundamental hurdles in the adoption of new technologies.
While \emph{trust} embodies several related principles, including explainability, dependability, robustness, etc., we loosely define a trustworthy technology as one that acts in the best selfish interest of its users.
The importance of trust has led \emph{game theory} to become a major thrust in contemporary control research~\citep{bacsar1998dynamic,fox2013population,grammatico2015decentralized,paccagnan2022utility}.
On the other hand, the importance of \emph{fairness} is widely acknowledged, yet this intuitive notion is difficult to define formally and incorporate in control objectives.
An emerging body of literature has thus aimed at incorporating fairness in control~\citep{jalota2021efficiency,villa2023fair,bang2024mobility,annaswamy2024equitable,elokda2024carma,shilov2025welfare,hall2025limits}.

This paper contributes to this emerging vision a novel roadmap to achieve \emph{trustworthy, fair, and efficient} resource allocations in socio-technical control settings, enabled by the recent development of \emph{karma economies}~\citep{elokda2024self,elokda2024carma}.
The concept of karma is illustrated in the following running example, adopted from~\cite{elokda2024carma}, which will be frequently revisited throughout the paper.

\begin{figure}[t]
    \centering
    \includegraphics[width=\textwidth]{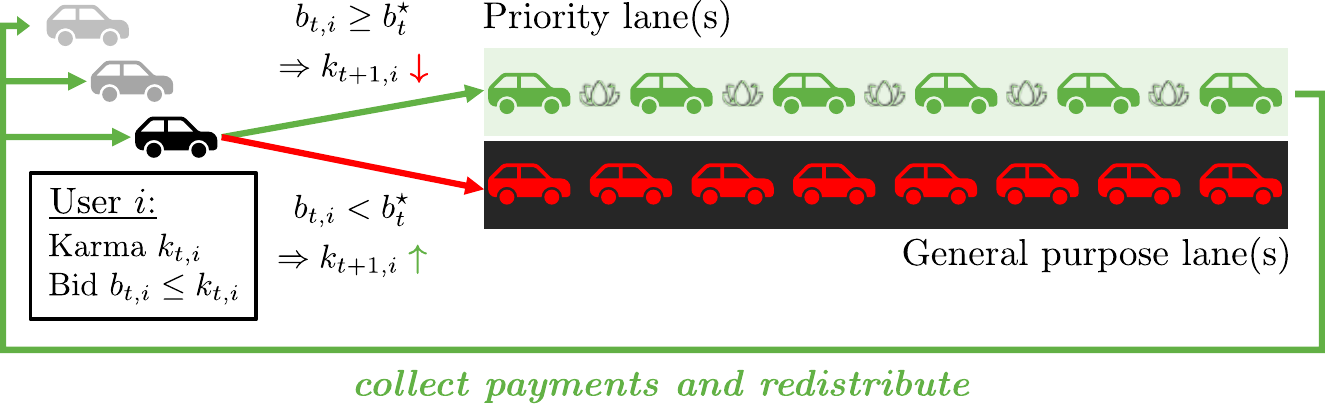}
    \caption{Illustration of a karma economy for the running example of allocating highway priority lanes.}
    \label{fig:running-example-karma}
\end{figure}
    
\begin{mdframed}[backgroundcolor=green!10]
\begin{example}[Allocation of Highway Priority Lanes]
\label{ex:running-example-w-karma}
    Figure~\ref{fig:running-example-karma} illustrates a common control measure aimed at reducing congestion in an arterial highway.
    The highway is divided into controlled \emph{priority lanes} and uncontrolled \emph{general purpose lanes}.
    Automation can be deployed to manage access to the priority lanes in real-time, and a control policy is needed to decide which cars should be given access.
    A karma economy can be used for this purpose, and it allocates the priority lanes repeatedly and indefinitely.
    Each user is endowed with non-tradable \emph{karma tokens} which are used to bid for access to the priority lanes on each day.
    The highest bidders, up to the free-flow capacity of the priority lanes, are granted entry; while all other users must use the general purpose lanes. 
    Those who enter the priority lanes then pay karma: in different variations of the mechanism, users can either pay their bid, or the \emph{uniform clearing bid}, denoted $b^\star$, which equals the lowest winning bid.
    Subsequently, at the end of each day, the total karma payment gets \emph{redistributed} to the users.
    Therefore, on each day, those who gain priority \emph{lose karma}; and those who lose priority \emph{gain karma} with which they can gain priority in future days.
    In short, the users are not only bidding against each other but \emph{playing against their future selves}.
\end{example}
\end{mdframed}

This example demonstrates how a karma economy leverages the repetitive nature of common socio-technical resource allocations to devise novel non-monetary, dynamic incentives.
The karma tokens budget individual resource consumption over time, enabling users to trade-off immediate versus future access to resources, or \emph{play against their future selves}.
This leads to \emph{efficient} allocations to the users with highest temporal \emph{urgency} for resources, according to the users' private urgency dynamics; and this efficiency is attained in a \emph{trustworthy} manner since it is in the best selfish interest of low urgency users to yield resources in favor for when it is their own turn to have high urgency.
Meanwhile, the resulting allocations are \emph{fair}, in a specific and rigorous sense formalized herein, because they provide users with equal opportunities to access resources over time.


To rigorously motivate karma economies, this paper develops four central \emph{pillars} that guide our vision for trustworthy, fair, and efficient socio-technical control.
A synopsis of these pillars, which are rooted in the economics literature and elaborated in Section~\ref{sec:pillars}, is provided here:
\begin{enumerate}
    \item \textbf{Single-shot and static solutions are limited:} When resource allocations are considered as \emph{single-shot and static}, it is generally impossible to jointly achieve trustworthy and efficient allocations;

    \item \textbf{Money does not always cut it:} The typical avenue to escape this impossibility is monetary mechanisms.
    But the assumption that money is fair to use in socio-technical contexts is highly subjective and controversial, as historically evidenced by public resistance to monetary control spanning multiple engineering applications;

    \item \textbf{Embrace the dynamics:} In contrast, when resource allocations are considered as \emph{repeated and dynamic}, new opportunities to jointly achieve trustworthy, fair, and efficient allocations present themselves.
    By budgeting resource consumption over time, as implemented by \emph{karma}, resource allocations can be tailored to the users' private urgency in an endogenous manner, i.e., without relying on external resources such as money;

    \item \textbf{A future built on Nash welfare:} However, in the definition of \emph{fairness and efficiency} that a dynamic allocation is ought to attain, one needs to carefully consider if the urgency of different users is \emph{interpersonally comparable}.
    Our last pillar advocates for \emph{Nash welfare}, a relatively understudied social welfare function with strong foundations in social choice theory and bargaining theory formalizing common intuition on what ought to be \emph{jointly fair and efficient}.
\end{enumerate}

With these pillars in place, we next proceed with formalizing the problem of allocating repeated resources dynamically in a manner that maximizes \emph{long-run Nash welfare}. First, in Section~\ref{sec:nash-welfare-allocation}, we take a centralized view on the problem to provide structural insights on \emph{Maximum Long-run Nash Welfare (MLNW)} solutions with examples.
Second, in Section~\ref{sec:karma} we formally introduce karma economies and their game-theoretical solution concept of \emph{Karma Equilibrium (KE)}; and show that KE coincide with MLNW under natural conditions.
This establishes the core building block to operationalize our vision for trustworthy, fair, and efficient socio-technical control: long-run Nash welfare jointly formalizes fairness and efficiency, and it is maximized in a trustworthy manner at the game-theoretical equilibrium of the karma economy.
Section~\ref{sec:outlook-multi-karma} thus lays out a possible future smart city re-imagined with karma.
The closed nature of the karma economy allows it to control the \emph{scope of fairness and efficiency}: it provides unprecedented flexibility to decide at what level(s) long-run Nash welfare should be maximized, and correspondingly, whether different karma economies should be coupled or kept separate.
Finally, Section~\ref{sec:conclusion} concludes the paper, pointing to several promising directions for future research on dynamic, non-monetary socio-technical control schemes.




\paragraph{Notation}
Throughout the paper, we denote scalar quantities in lightface $x$, while vectors, matrices, and higher dimensional tensors are denoted in boldface $\bm{x}$.
For example, a three-dimensional tensor is denoted as $\bm{x}=\left(x_{i,j,k}\right)_{i \in \I, j \in \J, k \in \K}$, with index sets $\I, \J, \K \subseteq \Nat$; for a particular $i \in \I$, $\bm{x_i} = \left(x_{i,j,k}\right)_{j \in \J, k \in \K}$ is the \emph{sub-matrix} corresponding to $i$; and similarly for $(i,j) \in \I \times \J$, $\bm{x_{i,j}}=\left(x_{i,j,k}\right)_{k \in \K}$ is the \emph{sub-vector} corresponding to $(i,j)$.
When $\bm{x} = \left(x_i\right)_{i \in \I}$ is a vector of quantities belonging to multiple users, for an ego user $i$, it is sometimes written as $\bm{x} = \left(x_i, \bm{x_{-i}}\right)$, where $\bm{x_{-i}} = \left(x_{i'}\right)_{i' \neq i}$ is the sub-vector for users other than $i$.
For a set $\X$, $\Delta(\X)$ denotes the space of probability measures over $\X$.

%% file: sections/pillars.tex
\section{Pillars of Trustworthy, Fair, and Efficient Socio-Technical Control}
\label{sec:pillars}





\subsection{Pillar 1: Single-shot and static solutions are limited}
\label{sec:static-no-good}

Socio-technical resource allocation problems are often studied as \emph{single-shot and static}~\citep{chremos2024mechanism,doostmohammadian2025survey}, despite being frequently repeated.
While many prior works consider \emph{microscopic system dynamics}, such as dynamic congestion models~\citep{de2011dynamic} and electric vehicle charging dynamics~\citep{parise2014mean}, they are static on a macroscopic time-scale: the fact that resources must be allocated multiple times over several periods or days is not modelled.

\begin{mdframed}[backgroundcolor=green!10]
\begin{example}(Single-shot Allocation of Highway Priority Lanes)
\label{ex:running-example-static}
Let us revisit our running example of allocating highway priority lanes, but for a single day only.
How can a real-time controller decide whom to grant and whom to deny access to the priority lanes?
A candidate controller could simply allocate \emph{random} users to the priority lanes up to their free-flow capacity.
This simple controller is ``efficient'' from a system-level perspective: as long as the priority lanes are fully utilized, optimal aggregate congestion reduction is achieved. However, it is clearly inefficient from the users' perspective, as it is agnostic to the their private urgency.
\end{example}
\end{mdframed}

It turns out, however, that there is not much that can be done to tailor the allocations to the private users' urgency, as the celebrated impossibility theorems of social choice theory assert~\citep{arrow1950difficulty,gibbard1973manipulation,satterthwaite1975strategy,hylland1980strategy}.
We recall the theorem due to~\cite{hylland1980strategy} which considers more general cardinal preferences and non-deterministic allocation rules.
Stating the theorem requires some preliminary notation.
Let $\X$ denote a set of allocations with at least three elements, e.g., which of the users enter the priority lanes, and let $\N=\{1,\dots,n\}$ be the index set of the users.
Each user $i \in \N$ has a private individual reward function over the allocations $r_i : \X \rightarrow \Real$, $r_i \in \R_i$, and $\bm{r} = \left(r_i\right)_{i\in\N} \in \R = \prod_{i \in \N} \R_i$ is the stacked profile of reward functions.
An \emph{allocation rule} $f : \R \rightarrow \Delta(\X)$ takes as input a profile of \emph{reported} reward functions $\bm{\tilde r} = \left(\tilde r_i\right)_{i\in\N} \in \R$ and outputs a probability distribution over $\X$ according to which an allocation is randomly chosen.
An allocation rule $f$ is said to be \emph{strategy-proof} if under $f$, it is a dominant strategy for each user to report its reward function truthfully.
An allocation rule $f$ is said to be \emph{ex-post Pareto efficient} if for all $\bm{r} \in \R$, all allocations in the support of $f(\bm{r})$ are \emph{Pareto efficient}, i.e., they are not dominated by any allocation that makes no users worse off and at least one user better off according to $\bm{r}$.
Notice that Pareto efficiency is the most primitive notion of efficiency one could hope for: in Example~\ref{ex:running-example-static}, for instance, any allocation that fully utilizes the priority lanes is Pareto efficient.

\begin{mdframed}[backgroundcolor=orange!10]
\begin{theorem}[\cite{hylland1980strategy}, Theorem~1]
\label{thm:static-impossibility}
    If an allocation rule $f$ is strategy-proof and ex-post Pareto efficient, then $f$ is a random dictatorship.
\end{theorem}
\end{mdframed}

A \emph{random dictatorship} can be described as follows:
choose a user randomly according to some probability distribution and implement its most preferred allocation; if the user has multiple most preferred allocations, choose a second user randomly and implement its most preferred allocation among those pre-selected by the first user; and so on.
Theorem~\ref{thm:static-impossibility} thus establishes an impossibility of jointly achieving \emph{trust} and \emph{efficiency}.
The intuition should be clear in light of Example~\ref{ex:running-example-static}:
any attempt to tailor the allocation to the private users' urgency will cause them to report the maximum urgency possible.
In fact, the simple random controller is precisely a ``random dictatorship'' in the sense of Theorem~\ref{thm:static-impossibility}.

\subsection{Pillar 2: Money does not always cut it}

Theorem~\ref{thm:static-impossibility} does not make any assumptions on the domain of reward functions $\R$ that can be any real-valued functions.
By assuming the existence of money, which has the same meaning to all agents (i.e., allows to make \emph{interpersonal comparisons}) and enters the reward functions in a \emph{quasi-linear} fashion~\citep{nisan2007introduction}, it becomes possible to design strategy-proof, Pareto efficient, and non-dictatorial rules, including the celebrated Vickrey-Clarke-Groves (VCG) mechanism~\citep{vickrey1961counterspeculation,clarke1971multipart,groves1973incentives}.
Accordingly, lots of socio-technical control research has focused on optimal monetary pricing policies~\citep{paschalidis2000congestion,mohsenian2010optimal,parise2014mean,barrera2014dynamic,gharesifard2015price,paccagnan2022utility,chremos2024mechanism}.
In Example~\ref{ex:running-example-static}, a controller could run real-time VCG auctions to allocate the lanes, or charge the theoretically optimal Vickrey toll~\citep{vickrey1969congestion} in real-time.
In an ideal world in which money is unequivocally fair, such a monetary scheme would allocate the priority lanes to the users with highest urgency, because they will have the highest willingness to pay.
But the world is far from ideal, and monetary solutions inherit income and wealth inequalities that compromise their fairness.
In what follows, we recall three historical examples that demonstrate that the use of money has been deemed unsatisfactory in many socio-technical control domains.

\begin{mdframed}[backgroundcolor=green!10]
\paragraph{\textbf{The case for net-neutral bandwidth allocation}}
The internet runs on the Transmission Control Protocol (TCP) and its celebrated \emph{additive-increase / multiplicative-decrease (AIMD)} distributed real-time congestion controller~\citep{chiu1989analysis}.
Each internet host follows a standard protocol in which it increases its sending rate additively if no congestion is detected, and decreases the rate multiplicatively when congestion is detected.
The network does not accommodate hosts that do not follow the standard protocol and their packets are dropped.
When all hosts follow the standard protocol, each of their sending rates is guaranteed to converge to the \emph{fair share} of the network capacity when congestion occurs\footnote{In single-link networks, the fair share is the link capacity divided by the number of hosts. In more general networks, different variants of AIMD lead to different notions of fair share, e.g., proportional~\citep{kelly1998rate} or max-min~\citep{mo2000fair}, but they are essentially similar.}.

One could envision an alternative controller that makes the internet more efficient by giving priority to more urgent flows.
To truthfully reveal the private urgency, a real-time bandwidth pricing scheme could be adopted.
This would essentially constitute a form of \emph{price discrimination} that led to a fierce debate in the United States since the 1990s, when it was suggested to deregulate the internet and let Internet Service Providers (ISPs) charge content hosts differently for available bandwidth~\citep{hahn2006economics}.
The debate culminated in a strong public sentiment for \emph{net-neutrality}: protecting the internet's integrity as a free and open resource that plays a crucial role in democratizing our modern societies~\citep{obama2016net}.
\end{mdframed}

\begin{mdframed}[backgroundcolor=green!10]
\paragraph{\textbf{73 years of debate on traffic congestion pricing}}
\label{sec:congestion-pricing}
Despite of traffic congestion being a severe societal problem in most of the world's major cities, leading to loss of time, air pollution, and poor quality of life, only a handful of cities have successfully adopted congestion pricing\footnote{At the time of writing, those are Singapore, London, Stockholm, Milan, Gothenburg, Oslo, Manhattan, and several other smaller cities and old town centers.}.
A typical congestion pricing policy charges drivers entering the city center a time-varying toll aimed at mimicking the real-time optimal price that would keep the city free of congestion.
The case of Manhattan, NYC is particularly interesting, as it resisted congestion pricing for almost 73 years despite being one of the most congested cities in the world.
Namely, William Vickrey, widely regarded as the father of congestion pricing~\citep{vickrey1969congestion}, originally conceptualized it for the NYC subway system in 1952, while the Manhattan policy finally came into effect in early 2025.
The long debate featured several rejected proposals (e.g., in 2008 and 2015) as well as protests and lawsuits against the policy, due to concerns that it would hurt marginalized communities that rely on driving to Manhattan from the boroughs~\citep{burkett2024teachers}, small business owners~\citep{trapani2023gottheimer}, and others~\citep{dias2023newyork}.
As a consequence of this controversy, the final policy was implemented in January 2025 at a peak toll price of \$9, instead of the originally proposed \$15, which was already a conservative approximation of the theoretically optimal price to eliminate congestion.

Notice that, other than tolling entry into a whole city, there are other real-world implementations of congestion pricing on smaller scales, notably the highway express lanes in California and other US states that largely resemble our running example (cf. Figure~\ref{fig:running-example-karma}).
These so-called High Occupancy/Toll (HOT) lanes have faced less controversy than tolling a whole city, or even the whole highway, since users at least have access to the free general purpose lanes~\citep{decorla2008income}.
Nonetheless, practical limitations on how high the toll can be persist, and HOT lanes have been nicknamed ``Lexus lanes'' due to public perception that they are ``built for the rich''~\citep{patterson2008lexus}.
\end{mdframed}

\begin{mdframed}[backgroundcolor=green!10]
\paragraph{\textbf{45 years of electricity demand response}}
Electricity is a socio-technical resource that has been facilitated by monetary markets for many years;
however, present electricity markets remain largely invisible to the end-user.
Most countries rely on large distribution operators to aggregate real-times prices over time and deliver constant or course time-of-day prices to the end-user~\citep{irena2019time}.
The idea of letting end-users participate actively in the electricity market, by either exposing them to real-time prices or providing them explicit incentives to reduce consumption during peak times~\citep{grvzanic2022prosumers}, has been present in a relatively mature form already since 1980, cf. \cite{schweppe1980homeostatic}, and classically referred to as \emph{demand side management}~\citep{gellings1985concept} or \emph{demand response} (DR)~\citep{daryanian1989optimal}.
However, thus far DR schemes have had limited effect in practice, primarily because the variability in real-time prices is too low for end-users to react in conventional power grids~\citep{grvzanic2022prosumers}.
DR schemes are expected to play a more pronounced role in future power grids due to the variability of renewable generation~\citep{eu2019directive}; but the effective operation of these schemes necessitates dramatic price discrimination that is likely to lead to public dismay.
\end{mdframed}

\subsection{Pillar 3: Embrace the dynamics}
\label{sec:pillar-dynamics}

In light of Theorem~\ref{thm:static-impossibility}, it may seem hopeless to design smart socio-technical controllers that tailor to user needs without the use of money.
Theorem~\ref{thm:static-impossibility} also provides an interpretation why most previous policy proposals have been largely polarized: either split resources equally with no regard for individual needs (e.g., net-neutrality, free or constant usage fee roads, or constant electricity prices); or adopt monetary control schemes (e.g., internet bandwidth market, congestion pricing, or demand response).
But Theorem~\ref{thm:static-impossibility} applies only to single-shot and static allocation problems, whereas many socio-technical resource allocation tasks are largely repetitive and dynamic, as the following simple extension of Example~\ref{ex:running-example-static} demonstrates.

\begin{figure}[!h]
\centering
\includegraphics[width=\textwidth]{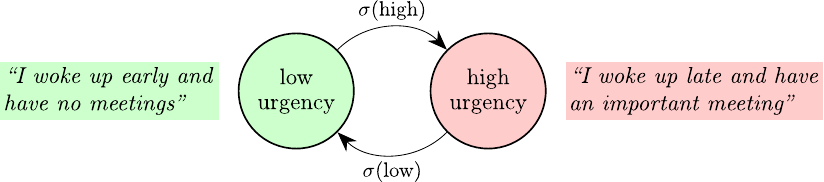}
\caption{Illustration of the concept of \emph{dynamic urgency} over several time-periods or days, which follows an i.i.d. process $\bm{\sigma}$. Self-loops are omitted for visual clarity.}
\label{fig:urgency-markov-chain}
\end{figure}
\begin{mdframed}[backgroundcolor=green!10]
\begin{example}[Two-shot Allocation of Highway Priority Lanes]
\label{ex:running-example-2shot}
    Consider that the priority lanes are not allocated once, but on two subsequent days.
    The private individual urgency for each day follows an exogenous independent and identically distributed (i.i.d.) process $\bm{\sigma}$, as depicted in Figure~\ref{fig:urgency-markov-chain}.
    Suppose that the priority lanes' free-flow capacity equals half of the travel demand, and that each user follows an  urgency process with $50\%-50\%$ chance of having low urgency ($u_\textup l$) or high urgency ($u_\textup h$) on any given day.
    Consider the following simple scheme to allocate the priority lanes.
    Each user gets one token with which it can enter the priority lanes on one of the two days.
    Ties, i.e., when more users than the free flow capacity `bid' their tokens, are settled randomly; but only those entering the priority lanes pay their token. 
    It can be shown that if $u_\textup h > 3 \, u_\textup l$, it is a dominant strategy for each user to play truthfully \emph{on the first day}, i.e., bid its token if and only if it has high urgency\footnote{The bound $u_\textup h > 3 \, u_\textup l$ corresponds to the corner case in which no one bids on the first day, and subsequently everyone bids on the second day. If $u_\textup l < u_\textup h \leq 3 \, u_\textup l$, bidding truthfully on the first day is not a dominant strategy, but nonetheless constitutes a (Bayesian) Nash equilibrium when all users follow it.}.
\end{example}
\end{mdframed}

Example~\ref{ex:running-example-2shot} demonstrates
that \emph{repetition, coupled with dynamics that link the allocations across time}, opens up new possibilities with respect to the single-shot and static setting. This is formalized by a recent but already seminal result by~\cite{jackson2007overcoming}.
Stating this result requires extending the notation introduced in Section~\ref{sec:static-no-good} to repeated and dynamic settings.
A finite set $\X \subset \Nat$ is repeatedly allocated from over $T \in \Nat$ time-steps.
In each time-step $t \in \T = \{1,\dots,T\}$, the private individual reward functions $r_{t,i}$ are drawn independently across time and users from a finite set $\tilde \R_i \subset \R_i$ according to publicly known probability distributions $\bm{\sigma_i} \in \Delta(\tilde \R_i)$, with $\bm{\sigma}$ denoting the corresponding joint distribution.
Let $f$ be a single-shot allocation rule as introduced in Section~\ref{sec:static-no-good}, and let $\bar r_i^f = \E_{\bm r \sim \bm{\sigma}, x \sim f(\bm r)}\left[r_i(x)\right]$ denote user $i$'s expected single-shot reward under $f$ (if all users report their rewards truthfully).
Allocation rule $f$ is said to be \emph{ex-ante Pareto efficient} if there is no dominating $f'$ for which $\bar r_i^{f'} \geq \bar r_i^f$ for all $i \in \N$ with strict inequality for at least one $i$.
Consider the following \emph{linking mechanism} aimed at implementing $f$ in a truth-revealing manner when repeated over $T$.
Each user $i$ receives a budget to report each $r_i \in \tilde \R_i$ exactly $\sigma_i(r_i) \, T$ times\footnote{For the sake of simplicity we assume here that $\sigma_i(r_i) \, T$ is integer for all $i \in \N$, $r_i \in \tilde \R_i$; c.f.~\cite{jackson2007overcoming} for the general treatment.}, and the allocation in each time-step $t$ is chosen according to $f$ evaluated at the present reports $\bm{\tilde r_t}$.
Let $\bm{\pi_i}=\left(\bm{\pi_{t,i}}\right)_{t \in \T}$ denote user $i$'s policy under this mechanism that specifies a probability distribution over its reports at each time-step (as a function of its true reward, the history of reports and allocations, etc.), and $\bm{\pi}=\left(\bm{\pi_i}\right)_{i \in \N}$ the policy profile of all users.
When policy profile $\bm{\pi}$ is followed, each user experiences an expected average reward of $\bar r_i(\bm{\pi}) = \cfrac{1}{T} \, \E\limits_{\substack{\bm{\tilde r_t} \sim \bm{\pi_t} \\ x_t \sim f(\bm{\tilde r_t}) \\ t \in \T}} \left[\sum\limits_{t \in \T} r_{t,i}(x_t)\right]$.
A policy profile $\bm{\pi^\star}$ is an \emph{$\epsilon$-approximate Bayesian Nash equilibrium} if no user can unilaterally improve its expected average reward by more than $\epsilon$, i.e., if $\bar r_i\left(\bm{\pi_i^\star},\bm{\pi^\star_{-i}}\right) \geq \bar r_i\left(\bm{\pi_i},\bm{\pi^\star_{-i}}\right) - \epsilon$ for all users $i \in \N$ and policies $\bm{\pi_i}$.
A policy profile in which all users report their rewards \emph{as truthfully as possible}\footnote{In the ex-post realization of rewards, it may not be possible to report them truthfully in every time-step under the linking mechanism, and the reports are as truthful as possible if they are truthful except for a minimal number of time-steps; cf. \cite{jackson2007overcoming} for the formal definition.} is denoted by $\bm{\pi^{\textup{true}}}$.

\begin{mdframed}[backgroundcolor=orange!10]
\begin{theorem}[\cite{jackson2007overcoming}, Corollary~2]
    \label{thm:dynamic-possibility}
    Let $f$ be an ex-ante Pareto efficient allocation rule.
    For any $\epsilon > 0$, there exists $\bar T \in \Nat$ such that for all $T \geq \bar T$, $\bm{\pi^\textup{true}}$ is an $\epsilon$-approximate Bayesian Nash equilibrium of the linking mechanism implementing $f$.
\end{theorem}  
\end{mdframed}

Theorem~\ref{thm:dynamic-possibility} essentially shows that \emph{dynamics enable escaping the static impossibility of jointly Pareto efficient, truthful, and non-dictatorial allocation rules}, as established in Theorem~\ref{thm:static-impossibility}.
To make this claim rigorous, some subtleties should be pointed out between both theorems.
The first subtlety is that Theorem~\ref{thm:static-impossibility} concerns \emph{ex-post} Pareto efficiency while Theorem~\ref{thm:dynamic-possibility} concerns \emph{ex-ante} Pareto efficiency; however, the ex-ante notion is stronger and leads to even less possibilities in the single-shot setting~\citep[Theorem~2]{hylland1980strategy}\footnote{In randomized allocations, \emph{``ex-ante''} refers to expected quantities before randomization takes place, and \emph{``ex-post''} to realized quantities after randomization. The only possibility to attain ex-ante Pareto efficiency and strategy-proofness in the single-shot setting is a (non-random) dictatorship.}.
The second subtlety is that Theorem~\ref{thm:static-impossibility} concerns \emph{strategy-proofness}, i.e., truthful reporting is a dominant strategy; while Theorem~\ref{thm:dynamic-possibility} concerns so-called \emph{Bayesian Nash incentive compatibility}, i.e., truthful reporting is an ($\epsilon$-approximate) Bayesian Nash equilibrium, which is weaker than strategy-proofness.
However, it is generally believed, though difficult to formally prove, that relaxing strategy-proofness to Bayesian Nash incentive compatibility does not open significant possibilities in the single-shot setting.
Some impossibility results for Bayesian Nash incentive compatibility have also been established recently~\citep{ehlers2020continuity,kikuchi2024general}.

While Theorem~\ref{thm:dynamic-possibility} provides hope that trustworthy, fair and efficient \emph{dynamic} resource allocation schemes are feasible, the linking mechanism used in the theorem is not well-suited for socio-technical control.
To summarize the shortcomings of this mechanism in control-theoretic terms, it is `open-loop.'
Namely, it requires precise knowledge of the reward function distributions $\bm{\sigma_i}$ to issue the correct amount of tokens.
Moreover, it is not suitable for allocations that are repeated indefinitely, as is the case in most socio-technical control contexts, since it would require issuing infinite tokens.
As discussed in Section~\ref{sec:related-mechanisms}, these shortcomings are addressed by our proposed `closed-loop' counter-part, that is, karma economies.


\subsection{Pillar 4: A future built on Nash welfare}
\label{sec:Nash-welfare}

\begin{figure}[!tb]
    \centering
    \begin{subfigure}[b]{0.48\textwidth}
        \centering
        \includegraphics[width=0.49\textwidth]{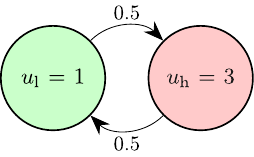}
        \hfil
        \includegraphics[width=0.49\textwidth]{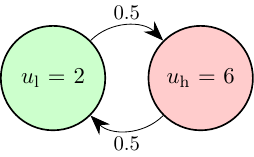}
        \caption{Same dynamics, different scale.}
        \label{fig:same-dynamics-different-scale}
    \end{subfigure}
    \hfill
    \begin{subfigure}[b]{0.48\textwidth}
        \centering
        \includegraphics[width=0.49\textwidth]{figures/urgency-markov-chain-example.pdf}
        \hfil
        \includegraphics[width=0.49\textwidth]{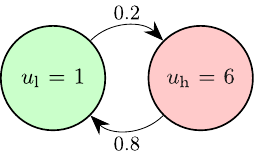}
        \caption{Same scale, different dynamics.}
        \label{fig:same-scale-different-dynamics}
    \end{subfigure}

\caption{Examples of private, heterogeneous i.i.d. urgency processes.
`Same scale' means that the average urgency is the same; `same dynamics' means that the frequency of urgency levels is the same (up to scaling).
How can urgency of different processes be compared to determine priorities?}
\label{fig:heterogeneous-urgency-chains}
\end{figure}

Consider that users have private and heterogeneous i.i.d. urgency processes, examples of which are shown in Figure~\ref{fig:heterogeneous-urgency-chains}.
If urgency quantifies the personal need for resources, can we assume that these needs can be expressed on a universal scale and compared across users with different processes?
It turns out that this question of \emph{interpersonal comparabilty}~\citep{roberts1980interpersonal,shilov2025welfare} has important consequences for all three overarching societal factors of trust, fairness and efficiency.

For \emph{trust},
consider the example of Figure~\ref{fig:same-dynamics-different-scale}, which shows two identical urgency processes up to scaling of the urgency levels.
If a dynamic mechanism distinguishes between these two processes, i.e., grants priorities based on absolute urgency, users will have an incentive to inflate their urgency scale.
One should thus expect any truthful dynamic mechanism to be \emph{scale-invariant}.

For \emph{fairness} and \emph{efficiency}, which we discuss jointly under the more general term of \emph{social welfare}, one should be mindful of the relevant information the urgency processes are meant to portray.
If there is no reason to grant time-persistent priorities to some users over others, the urgency scale should once again play no role.
Figure~\ref{fig:same-scale-different-dynamics} shows an example of two urgency processes with the `same scale', in the sense that they have the same long-run average urgency, but different dynamic natures.
Despite having `same scale', it is not straightforward to compare urgency levels of different processes, since these levels are experienced at different frequencies.
Nonetheless, there is room to unequivocally promote both fairness and efficiency \emph{without} making interpersonal comparisons of urgency.
For example, it is natural to prioritize urgency $6$ in Figure~\ref{fig:same-scale-different-dynamics}-right over urgency $3$ in Figure~\ref{fig:same-scale-different-dynamics}-left, not because $6>3$ in an absolute sense, but because urgency $6$ occurs less frequently; and in turn, forms a more intense \emph{intrapersonal} deviation relative to other days than urgency $3$ does.


The above discussion can be summarized in requiring that urgency be \emph{interpersonally non-comparable}.
Fortunately, there is a social welfare function that elegantly encodes the possible efficiency gains under this assumption, that is, the \emph{Nash welfare function}, given by
\begin{mdframed}[backgroundcolor=red!10]
\begin{align}
    \textup{NW}\left(\bm{\chi}\right) = \prod_{i \in \N} \left(\bar r_i\left(\bm{\chi}\right) - \bar r_i\left(\bm{\chi^0}\right)\right)^{w_i}. \tag{NW} \label{eq:Nash-welfare}
\end{align}
\end{mdframed}
In~\eqref{eq:Nash-welfare}, $\bm{\chi} \in \Delta(\X^T)$ denotes a probabilistic allocation over all times $t \in \T$, $\bar r_i\left(\bm{\chi}\right)$ is the expected average reward of user $i \in \N$ under $\bm{\chi}$, and $\bm{\chi^0} \in \Delta(\X^T)$ is a \emph{benchmark or status-quo allocation} in which all users are worst-off.
In the running example of allocating priority lanes, the benchmark could be that no control measure is introduced, i.e., all lanes are general purpose, cf. Example~\ref{ex:running-example-full}.
Finally, $\bm{w} = \left(w_i\right)_{i \in \N} \in \Real_{>0}^n$ denote exogenous user-specific priority weights or \emph{access rights}.
Most commonly, $w_i = 1$ for all $i \in \N$ means that all users have equal rights; but there are cases in which unequal rights could be warranted, e.g., prioritizing emergency care givers over regular citizens in accessing fast lanes.

Nash welfare is a classical social welfare function that has been portrayed to \emph{`strike a good trade-off between fairness and efficiency'}~\citep{caragiannis2019unreasonable,bertsimas2012efficiency}.
This motivation can be misleading because `trading off' fairness and efficiency implicitly assumes interpersonal comparability\footnote{Namely, the sum of rewards that is most commonly used to define efficiency, as well as most existing fairness measures (e.g., minimum reward, Gini index, etc.), assume interpersonal comparability.}.
In what follows, we argue that Nash welfare \emph{jointly encodes fairness and efficiency}, rather than trades off between the two.

\begin{mdframed}
\paragraph{\textbf{The case for Nash welfare from social choice theory}}
Parallel to studying the possibilities and impossibilities of designing allocation rules ala Theorems~\ref{thm:static-impossibility}--\ref{thm:dynamic-possibility}, social choice theory investigates how to meaningfully aggregate individual preferences into \emph{social welfare functions}~\citep{roberts1980interpersonal}.
Examples of social welfare functions are the utilitarian sum of rewards, the egalitarian minimum reward, and Nash welfare.
\cite{shilov2025welfare} provides a comprehensive overview of results in this space tailored to a control audience, and we provide an informal summary of the basic elements here.
The choice of social welfare function is guided by the following set of basic properties or axioms that the function should satisfy:

\smallskip

\begin{itemize}
    \item \emph{Pareto principle (P):} If allocation $x$ Pareto dominates allocation $y$, then $x$ should attain higher social welfare than $y$;
    
    \item \emph{Independence of irrelevant alternatives (IIA):} The social ordering of two allocations $x$ and $y$, i.e., which allocation of the two attains higher social welfare, depends only on the users' rewards at $x$ and $y$ and not on the rewards of other allocations\footnote{(IIA) is the most often critiqued axiom of social choice theory. However, in addition to its basic premise, (IIA) has an important connection to \emph{strategy-proofness}: under additional mild conditions, an allocation rule is strategy-proof if and only if it maximizes a social welfare function satisfying (IIA)~\citep{satterthwaite1975strategy}.}.

    \item \emph{Comparability:} Assumptions on how to compare the rewards of different users range from the weakest requirement of \emph{ordinal non-comparability (ONC)}, which considers rewards as ordinal\footnote{\emph{Ordinal rewards} express rankings of allocations: they tell if an allocation is preferred over another, but not by how much. \emph{Cardinal rewards} additionally express preference intensities in relative terms: for any three allocations $x$, $y$, $z$, they quantify the reward difference between $x$ and $y$ relative to the difference between $x$ and $z$; cf.~\cite{shilov2025welfare} for the formal definitions.} and not interpersonally comparable; to the strongest requirement of \emph{perfect comparability (PC)}, which considers that rewards are perfectly comparable on a common absolute scale.
\end{itemize}

\smallskip

Depending on the adopted comparability requirement, there is either no possibility, a unique possibility, or a restricted set of possibilities to jointly satisfy (P) and (IIA)~\citep[Theorem~1]{shilov2025welfare}.
Remarkably, the \emph{weakest} comparability requirement for which it becomes possible to satisfy (P) and (IIA) is \emph{cardinal non-comparability (CNC)}, which postulates that rewards are cardinally \emph{intrapersonally} comparable but \emph{interpersonally} non-comparable.
Under (CNC), the \emph{unique} social welfare function satisfying (P) and (IIA) is the Nash welfare function~\eqref{eq:Nash-welfare}.
The worst-case benchmark allocation $\bm{x^0}$ is important for this result, namely, without such a benchmark it is impossible to satisfy (P), (IIA), and (CNC).
Intuitively, the benchmark circumvents this impossibility because it acts as a common point with respect to which relative \emph{intrapersonal} improvements can be compared.
In light of the preceding discussion to regard urgency of heterogeneous processes as interpesonally non-comparable, social choice theory thus provides strong foundations for adopting~\eqref{eq:Nash-welfare} as the social welfare function in dynamic resource allocation problems.
\end{mdframed}

\begin{mdframed}
\paragraph{\textbf{The case for Nash welfare from bargaining theory}}
A parallel and independent motivation for Nash welfare stems from \emph{bargaining theory}.
In fact, the name Nash welfare attributes its origins to the \emph{Nash bargaining solution}~\citep{nash1950bargaining}\footnote{The Nash bargaining solution is not to be confused by the famous Nash equilibrium, and in fact Nash welfare has little to nothing to do with the Nash equilibrium.}.
In the bargaining problem, two or more users must come to an agreement on an allocation $x$, and if negotiations fail, they must fall back on the status quo or default option $x^0$.
The bargaining problem is ill-defined because there are often many allocations that would benefit all users, but could differ significantly in the distribution of gains.
The Nash bargaining solution, which maximizes Nash welfare, is widely regarded as the fair outcome of the bargain: in the case of dividing monetary gains, it leads to splitting the surplus equally after compensating the users for the \emph{opportunity cost} of forfeiting the default option $x^0$.
Moreover, if the bargaining is performed sequentially or repeatedly, it has been shown that the Nash bargaining solution emerges as the unique equilibrium~\citep{rubinstein1982perfect,binmore2005natural}.
Based on this principle, it has been suggested that moral principles are in fact social contracts that have evolved to maximize Nash welfare~\citep{andre2022evolutionary}.

One could adopt a bargaining interpretation of socio-technical control problems.
Society must come to an agreement on what control scheme to adopt.
In the case of NYC congestion pricing, c.f. Section~\ref{sec:congestion-pricing}, the negotiations lasted 73 years; perhaps they would have concluded quicker if one targeted maximum Nash welfare.
\end{mdframed}


To summarize, we presented several motivations for adopting Nash welfare as a social welfare function that jointly encompasses fairness and efficiency in socio-technical control; yet it is has received relatively little attention in comparison to other functions (e.g., the sum or the minimum of rewards).
We postulate that this is partially because Nash welfare leads to trivial allocations in many single-shot allocation problems: due to its scale invariance property, it disregards many static heterogeneities and allocates resources randomly.
Our upcoming Section~\ref{sec:nash-welfare-allocation} is thus dedicated to the study of \emph{long-run Nash welfare} in infinitely repeated and dynamic contexts, in which it is remarkably possible to extract efficiency gains without violating fairness.
Following, in Section~\ref{sec:karma}, we show that long-run Nash welfare can be maximized in a trustworthy manner with karma economies.

%% file: sections/problem.tex
\section{Dynamic Resource Allocation under Maximum Long-Run Nash Welfare (MLNW)}
\label{sec:nash-welfare-allocation}

In this section we formally introduce \emph{dynamic resource allocation problems} (Section~\ref{sec:problem-formulation}) and then we take a centralized view on \emph{Maximum Long-Run Nash Welfare (MLNW)} solutions to highlight their fairness and efficiency properties (Section~\ref{sec:max-Nash-welfare}).

\subsection{Problem formulation}
\label{sec:problem-formulation}

Our problem formulation adapts and extends canonical single-shot resource allocation models~\citep{vazirani2007combinatorial,hylland1979efficient} to consider infinite repetition and dynamic urgency.
We consider the problem of repeatedly allocating resources $\M=\{1,\dots,m\}$ (indexed $j$) of capacities $\bm c = \left(c_j\right)_{j \in \M} \in \Nat_{>0}^m$ to users $\N=\{1,\dots,n\}$ over an infinite time.
The users have exogenous \emph{access rights} $\bm{w} = (w_i)_{i \in \N} \in \Real_{>0}^n$ representing long-run priorities in accessing resources.
The allocation at time-step $t \in \Nat$ is denoted by $\bm{x_t} \in \X = \left\{\bm x \in \{0,1\}^{n \times m}; \; \sum_{i \in \N} x_{i,j} \leq c_j, \; \forall j \in \M \right\}$, where $x_{t,i,j}=1$ (respectively, $x_{t,i,j}=0$) denotes that user $i$ is allocated (respectively, not allocated) resource $j$.
The sub-allocation to user $i$ is denoted by $\bm{x_{t,i}} = \left(x_{t,i,j}\right)_{j \in \M} \in \{0,1\}^m$.
In some settings, the resources are \emph{mutually exclusive}, i.e., each user can be allocated at most one resource at a time, for which it must also hold that $\sum_{j \in \M} x_{t,i,j} \leq 1$  for all $i \in \N$.

Each user $i \in \N$ derives a private \emph{elementary reward} $r_{i,j} \in \Real$ for receiving resource $j \in \M$, while not receiving resource $j$ yields an elementary reward of $r^0_{i,j} \in \Real$.
Notice that resources $j$ for which $r_{i,j} < r^0_{i,j}$ are actually \emph{bads} for user $i$, i.e., it prefers to not receive these resources.
We denote by $\N_j = \left\{i \in \N; \; r_{i,j} > r^0_{i,j}\right\} \subseteq \N$ the set of users that desire resource $j$, and make the following standing assumption.
\begin{assumption}[Competitive Setting]
\label{as:competition}
The following holds:
\begin{enumerate}[label=\ref{as:competition}.\arabic*]
    \item \label{as:competition-agent} Each user desires at least one resource, i.e., for all $i \in \N$, there exists $j \in \M$ such that $i \in \N_j$;
    \item \label{as:competition-resource} Each resource is contested, i.e., for all $j \in \M$, $c_j < \left\lvert\N_j\right\rvert$.
\end{enumerate}
\end{assumption}

Moreover, each user has a time-varying private \emph{urgency} $u_{t,i} \in \U_i=\{u_i^1,\dots,u_i^{q_i}\} \subset \Real_{> 0}$ that multiplies its elementary reward.
Accordingly, user $i$'s stage reward given urgency $u_{t,i}$ and allocation $\bm{x_t}$ is given by
\begin{multline*}
    r_i(u_{t,i},\bm{x_t}) = r_i(u_{t,i},\bm{x_{t,i}}) = u_{t,i} \sum_{j \in \M} \left(x_{t,i,j} \, r_{i,j} + \left(1 - x_{t,i,j}\right) r^0_{i,j} \right) \\
    = u_{t,i} \sum_{j \in \M} \left(x_{t,i,j} \left(r_{i,j} - r^0_i\right) + r^0_{i,j} \right).
\end{multline*}
The urgency $u_{t,i}$ of each user follows a private, exogenous stochastic process that is independent across users and time, denoted by $\bm{\sigma_i} \in \Delta(\U_i)$ and satisfying $\sigma_i(u_i) > 0$ for all $u_i \in \U_i$.
The dynamic resource allocation problem is thus specified by the tuple $\DRA=\left\{\N,\M,\bm{w},\bm{c},\bm{r},\bm{r^0},\bm{\U}, \bm{\sigma}\right\}$, where $\bm{r}=\left(r_{i,j}\right)_{i\in\N,j\in\M}$, $\bm{r^0}=\left(r^0_{i,j}\right)_{i\in\N,j\in\M}$, $\bm{\U}=\left(\U_i\right)_{i\in\N}$, and $\bm{\sigma}=\left(\bm{\sigma_i}\right)_{i\in\N}$.
We illustrate this notation on the following extension of our running example.

\begin{mdframed}[backgroundcolor=green!10]
\begin{example}[Allocation of Highway Priority Lanes over Morning Rush Hour]
\label{ex:running-example-full}
Consider now that the highway priority lanes are allocated repeatedly \emph{over the morning rush hour}.
Following~\cite{elokda2024carma} that is based on the classical bottleneck model~\citep{vickrey1969congestion}, suppose that every morning, $n=9000$ single occupancy vehicle users must traverse a highway of total capacity $60$ vehicles per minute; thus the morning rush hour lasts $150$ minutes.
The highway is split into priority lanes of capacity $12$ vehicles per minute ($20\%$ of the total capacity), and general purpose lanes of capacity $48$ vehicles per minute.
Let us discretize the departure times into $15$-minute intervals, such that the rush hour lasts over $10$ departure intervals.
We are thus faced with the problem of repeatedly allocating $\M=\{1,\dots,10\}$ mutually exclusive resources, corresponding to the priority lanes at different departure intervals, whose capacities are $c_j = c = 180$ for all $j \in \M$, that is, the free-flow capacity of the priority lanes over one interval.
The departure intervals are not equally favorable to the users.
Namely, all users $i \in \N$ wish to depart at interval $j^*=9$ (e.g., corresponding to 9:00am), and experience a reward of $r_{i,j} = r_j = -\max\{j^* - j, \, 4 \left(j - j^*\right)\}$, i.e., they incur a proportionally larger discomfort the further they depart from $j^*$, and are more sensitive towards late than early departures.
In case a user is not allocated to the priority lanes at any interval, they must use the general purpose lanes, which gives a reward of $r^0_{i,j} = r^0=-8$ corresponding to the non-controlled equilibrium discomfort in those lanes\footnote{The non-controlled equilibrium discomfort of the general purpose lanes additionally takes into account delays due to the build-up of queues.
It coincides with the queue-free discomfort of departing in the earliest interval $j=1$, as well as the equilibrium discomfort if \emph{all} lanes are not controlled, cf.~\cite{elokda2024carma}.}.
Notice that Assumption~\ref{as:competition} holds in this setting with the exception of the earliest interval $j=1$ which gives the same reward as $r^0$.
Moreover, suppose that all users have equal access rights, i.e., $w_i = 1$ for all $i \in \N$.
\end{example}
\end{mdframed}

\subsection{The fairness and efficiency of Maximum Long-run Nash Welfare (MLNW)}
\label{sec:max-Nash-welfare}

To showcase the fairness and efficiency of MLNW in dynamic resource allocation problems, in this section, we take the viewpoint of an all-knowing centralized controller that can directly control the long-run probability of each user $i \in \N$ receiving resource $j \in \M$ when in urgency $u_i \in \U_i$, denoted by $\chi_{i,j}(u_i) \in [0,1]$.
We write the \emph{long-run allocation} to user $i$ as $\bm{\chi_i} = \left(\chi_{i,j}(u_i)\right)_{j \in \M, u_i \in \U_i} \in [0,1]^{m \times q_i}$, and $\bm{\chi} = \left(\bm{\chi_i}\right)_{i \in \N}$ is simply the \emph{long-run allocation}.
User $i$'s long-run expected average reward at $\bm{\chi}$ is thus given by
\begin{multline}
    \bar r_i(\bm{\chi}) = \bar r_i(\bm{\chi_i}) = \lim_{T \rightarrow \infty} \frac{1}{T} \, \E\limits_{\substack{u_{t,i} \sim \bm{\sigma_i} \\ \bm{x_{t,i}} \sim \bm{\chi_{i}}(u_{t,i}) \\ t \in \{1,\dots,T\}}} \left[\sum_{t=1}^T r_{t,i}(u_{t,i},\bm{x_{t,i}})\right] \\
    = \sum_{u_i \in \U_i} \sigma_i(u_i) \, u_i \sum_{j \in \M} \left(\chi_{i,j}(u_i) \left(r_{i,j} - r^0_{i,j}\right) + r^0_{i,j} \right).
\end{multline}

To complete the specification of the MLNW problem, we require user-specific access rights $\bm{w}$ as well as a worst-case benchmark allocation $\bm{\chi^0}$.
In what follows, we consider the natural benchmark of \emph{no allocation} $\bm{\chi^0}=\bm{0}$, and therefore $\bar r_i(\bm{\chi}) - \bar r_i(\bm{\chi^0}) := \bar r_i - \bar r_i^0 = \sum_{u_i \in \U_i} \sigma_i(u_i) \, u_i \sum_{j \in \M} \chi_{i,j}(u_i) \left(r_{i,j} - r^0_{i,j}\right)$.
This gives the following optimization problem

\begin{mdframed}[backgroundcolor=red!10]
\begin{minipage}{0.22\textwidth}\vspace{0pt}
    \textbf{MLNW Problem:}
\end{minipage}
\hfill
\begin{minipage}{0.75\textwidth}\vspace{0pt}
\begin{equation}
\begin{aligned}
    \max_{\bm \chi} &&& \sum_{i \in \N} w_i \log \left(\bar r_i - \bar r^0_i\right) \\
    \textup{subject to} &&& \sum_{i \in \N} \sum_{u_i \in \U_i} \sigma_i(u_i) \, \chi_{i,j}(u_i) \leq c_j, \; \forall j \in \M, \\
    &&& 0 \leq \chi_{i,j}(u_i) \leq 1, \; \forall i \in \N, j \in \M, u_i \in \U_i.
\end{aligned}
\tag{MLNW} \label{eq:max-Nash-welfare}
\end{equation}
\end{minipage}
\end{mdframed}

Notice that we use the logarithm of the Nash welfare function~\eqref{eq:Nash-welfare} in the objective (which is equivalent since the logarithm is a monotonic transformation), and moreover require that the resource capacity constraints are satisfied \emph{ex-ante}\footnote{The capacity constraints are satisfied \emph{ex-post} almost surely in the large population limit: assuming that users can be grouped into a finite number of types with identical urgency processes,
the ex-post urgency distribution of every time-step converges to $\bm{\sigma}$ almost surely.}.
In settings where the resources are mutually exclusive, it must additionally hold that $\sum_{j \in \M} \chi_{i,j}(u_i) \leq 1$ for all $i \in \N$, $u_i \in \U_i$.

Since Problem~\eqref{eq:max-Nash-welfare} is convex and Slater's constraint qualification holds\footnote{Under Assumption~\ref{as:competition}, a feasible interior point is obtained with $\chi_{i,j}(u_i) = \frac{c_j - \epsilon}{\left\lvert\N_j\right\rvert}$ for all $j \in \M$, $i \in \N_j$, $u_i \in \U_i$, and some small $\epsilon > 0$.}, the Karush-Kuhn-Tucker (KKT) conditions are necessary and sufficient for optimality.
Let $(\bm{\chi^*}, \bm{\lambda^*}, \bm{\eta^*})$ be the solution of the KKT system, i.e., $\bm{\chi^*}$ solves Problem~\eqref{eq:max-Nash-welfare} with optimal Lagrange multipliers $\bm{\lambda^*} \in \Real_{\geq 0}^m$ for the resource capacity constraints and $\bm{\eta^*_i} \in \Real_{\geq 0}^{m \times q_i}$, $\bm{\eta^*} =\left(\bm{\eta^*_i}\right)_{i \in \N}$ for constraints $\chi_{i,j}(u_i) \leq 1$.
Denote further user $i$'s long-run expected average reward at the optimum by $\left(\bar r_i - \bar r_i^0\right)^* := \bar r_i(\bm{\chi^*}) - \bar r_i^0$.
We have the following characterization of maximum Nash welfare solutions.

\begin{mdframed}[backgroundcolor=blue!10]
\begin{proposition}[Properties of MLNW]
    \label{prop:max-Nash-welfare}
    Let Assumption~\ref{as:competition} hold.
    Then MLNW solutions $(\bm{\chi^*}, \bm{\lambda^*}, \bm{\eta^*})$ satisfy, for all $i \in \N$, $j \in \M$, $u_i \in \U_i$:
    \begin{enumerate}[label=\ref{prop:max-Nash-welfare}.\arabic*]
        
        \item \label{prop:max-Nash-all-improve} $\left(\bar r_i - \bar r^0_i\right)^* > 0$ and is unique;

        \item \label{prop:max-Nash-scale-invariance} If two users $i$, $i'$ are symmetric up to scales $\alpha_i, \alpha_{i'} >0$\footnote{Two users $i$, $i'$ are called \emph{symmetric up to scales} $\alpha_i, \alpha_{i'} > 0$ if they share the same characteristics up to scaling of the urgency, i.e., $w_i = w_{i'}$, $\bm{r_i}=\bm{r_{i'}}$, $\bm{r^0_i}=\bm{r^0_{i'}}$, $\bm{\phi_i}=\bm{\phi_{i'}}$, and there exists a common urgency set $\U$ such that $\U_i = \alpha_i \, \U$ and $\U_{i'} = \alpha_{i'} \, \U$.}, then $\cfrac{\left(\bar r_i - \bar r^0_i\right)^*}{\alpha_i} = \cfrac{\left(\bar r_{i'} - \bar r^0_{i'}\right)^*}{\alpha_{i'}}$;

        \item \label{prop:max-Nash-high-urgency-first} $\chi^*_{i,j}(u_i) > 0 \Rightarrow \chi^*_{i,j}(u'_i) = 1$, for all $u'_i > u_i$;
        
        \item \label{prop:max-Nash-resources-used} $\lambda^*_j > 0$, and all resources are allocated at capacity;

        
        \item \label{prop:max-Nash-stationarity-inequality} $u_i \left(r_{i,j} - r^0_{i,j}\right) \leq \cfrac{\left(\bar r_i - \bar r^0_i\right)^*}{w_i} \left(\lambda^*_j + \cfrac{\eta^*_{i,j}(u_i)}{\sigma_i(u_i)} \right)$;

        \item \label{prop:max-Nash-stationarity-equality} $\chi^*_{i,j}(u_i) > 0 \Rightarrow u_i \left(r_{i,j} - r^0_{i,j}\right) = \cfrac{\left(\bar r_i - \bar r^0_i\right)^*}{w_i} \left(\lambda^*_j + \cfrac{\eta^*_{i,j}(u_i)}{\sigma_i(u_i)} \right)$;
        
        \item \label{prop:max-Nash-weight} $\sum\limits_{u_i \in \U_i} \sigma_i(u_i) \sum\limits_{j \in \M} \chi^*_{i,j}(u_i) \, \lambda^*_j + \sum\limits_{j \in \M} \sum\limits_{u_i \in \U_i} \eta^*_{i,j}(u_i) = w_i$.

    \end{enumerate}

    Moreover, if the resources are mutually exclusive, then Propositions~\ref{prop:max-Nash-all-improve}--\ref{prop:max-Nash-scale-invariance} hold, Propositions~\ref{prop:max-Nash-stationarity-inequality}--\ref{prop:max-Nash-weight} hold with non-resource specific multipliers $\eta^*_{i,j}(u_i)=\eta^*_i(u_i)$, and it holds for all $i \in \N$, $j \in \M$, $u_i \in \U_i$:
    \begin{enumerate}
        \item[\ref{prop:max-Nash-high-urgency-first}']  \label{prop:max-Nash-high-urgency-first-mutually-exclusive} $\chi^*_{i,j}(u_i) > 0 \Rightarrow \sum_{j' \in \M} \chi^*_{i,j'}(u'_i) = 1$, for all $u'_i > u_i$;
        
        \item[\ref{prop:max-Nash-resources-used}'] \label{prop:max-Nash-resources-used-mutually-exclusive} $\lambda^*_j = 0 \Rightarrow \sum_{j' \in \M} \chi^*_{i,j'}(u_i) = 1$, for all $i \in \N_j$, $u_i \in \U_i$.
    \end{enumerate}
\end{proposition}
\end{mdframed}

The proof of Proposition~\ref{prop:max-Nash-welfare} is included in~\ref{app:proof-max-Nash-welfare}.
Maximum Nash welfare solutions satisfy three important fairness and efficiency properties. 
First, they guarantee that all users \emph{strictly} benefit with respect to the benchmark $\bm{\chi^0}$, c.f. Proposition~\ref{prop:max-Nash-all-improve}.
Second, they satisfy the \emph{equal treatment of equals} principle in a scale-invariant manner, c.f. Proposition~\ref{prop:max-Nash-scale-invariance}.
Third, they satisfy a natural efficiency property of \emph{serving the highest urgency first}, cf. Proposition~\ref{prop:max-Nash-high-urgency-first}.
In addition, all resources are allocated at capacity,
and the optimal resource constraint multipliers $\bm{\lambda^*}$ are strictly positive, cf. Proposition~\ref{prop:max-Nash-resources-used}.
Notice that if for a user $i \in \N$, resource $j \in \M$, and urgency $u_i \in \U_i$, constraint $\chi^*_{i,j}(u_i) = 1$ binds with $\eta^*_{i,j}(u_i) > 0$, it must hold that $u_i \left(r_{i,j} - r^0_{i,j}\right) > \frac{\left(\bar r_i - \bar r^0_i\right)^*}{w_i} \, \lambda_j^*$, cf. Proposition~\ref{prop:max-Nash-stationarity-equality}.
If one interprets $\lambda^*_j$ as a \emph{shadow price} of resource $j$, this means that user $i$ in urgency $u_i$ would be willing to `pay' a higher price to get a larger share of resource $j$ if it spent more time in that urgency, i.e., if $\sigma_i(u_i)$ is increased.
With this same interpretation of $\bm{\lambda^*}$ as shadow prices, Proposition~\ref{prop:max-Nash-weight} reveals that the access rights $\bm{w}$ act as \emph{budget constraints on the long-run expected average shadow payment}, since $\sum_{u_i \in \U_i} \sigma_i(u_i) \sum_{j \in \M} \chi^*_{i,j}(u_i) \, \lambda^*_j \leq w_i$ for all users $i \in \N$, with equality if constraints $\chi^*_{i,j}(u_i) \leq 1$ do not bind for $i$, i.e., $\sum_{j \in \M} \sum_{u_i \in \U_i} \eta^*_{i,j}(u_i) = 0$. 

In settings where the resources are mutually exclusive, the maximum Nash welfare solution has similar structural properties with the exception that constraints $\sum_{j \in \M}\chi^*_{i,j}(u_i) \leq 1$ bind before constraints $\chi^*_{i,j}(u_i) \leq 1$, and therefore it holds that $\eta^*_{i,j}(u_i) = \eta^*_i(u_i)$ for all $i \in \N$, $j \in \M$, $u_i \in \U_i$.
Moreover, there could be some resources $j \in \M$ that are not fully allocated and/or have $\lambda^*_j = 0$. For these resources, it must hold that all users $i \in \N_j$ are guaranteed to receive a (potentially more desirable) resource in all urgency states $u_i \in \U_i$.
To complement the theoretical properties established in Proposition~\ref{prop:max-Nash-welfare}, we revisit our running example of allocating highway priority lanes.

\begin{figure}[!tb]
\centering
\begin{subfigure}{\textwidth}
    \centering
    \includegraphics[height=0.15\textwidth]{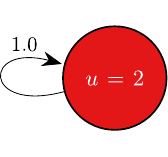}
    \hfil
    \includegraphics[height=0.15\textwidth]{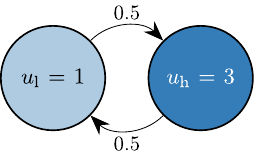}
    \hfil
    \includegraphics[height=0.15\textwidth]{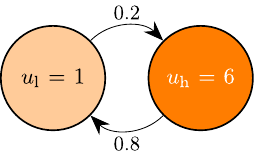}
    \hfil
    \includegraphics[height=0.15\textwidth]{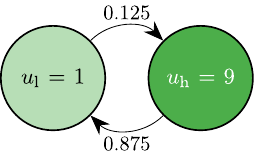}
    \caption{Heteregoneous urgency processes used in Example~\ref{ex:running-example-max-Nash-welfare}.}
    \label{fig:heterogeneous-urgency-example}
\end{subfigure}

\begin{subfigure}{\textwidth}
    \centering
    \includegraphics[width=0.38\textwidth]{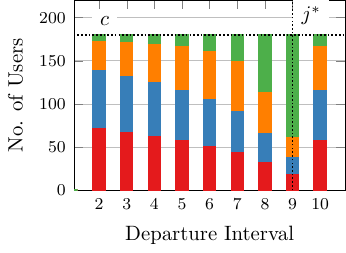}
    \hfil
    \includegraphics[width=0.38\textwidth]{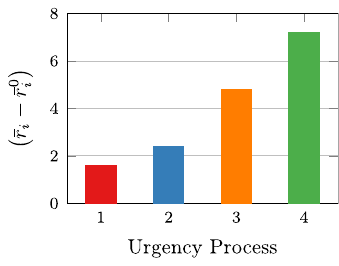}
    \caption{\textbf{Maximum Long-run Nash Welfare (MLNW) Solution}.}
    \label{fig:max-Nash-welfare-solution}
\end{subfigure}

\begin{subfigure}{\textwidth}
    \centering
    \includegraphics[width=0.38\textwidth]{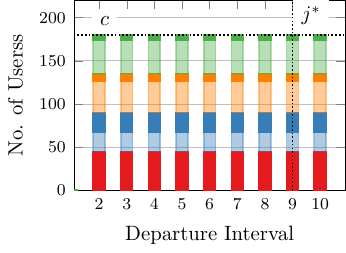}
    \hfil
    \includegraphics[width=0.38\textwidth]{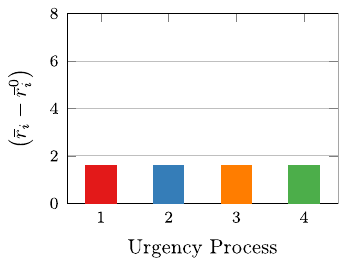}
    \caption{\emph{Single-shot} Maximum Nash Welfare Solution.}
    \label{fig:static-Nash-welfare-solution}
\end{subfigure}

\begin{subfigure}{\textwidth}
    \centering
    \includegraphics[width=0.38\textwidth]{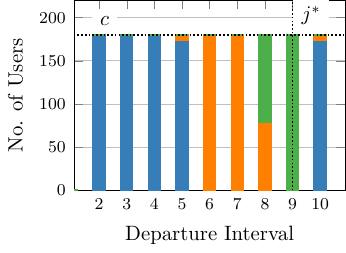}
    \hfil
    \includegraphics[width=0.38\textwidth]{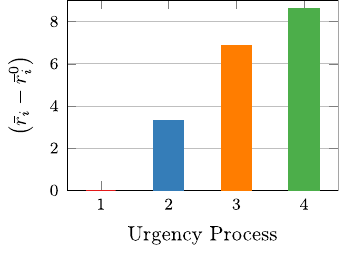}
    \caption{Utilitarian Solution.}
    \label{fig:utilitarian-solution}
\end{subfigure}

\caption{Priority lanes allocation over the morning rush hour under the heterogeneous urgency processes depicted in Figure~\ref{fig:heterogeneous-urgency-example}. Figures~\ref{fig:max-Nash-welfare-solution}--\ref{fig:utilitarian-solution} compare three solutions, showing the allocation of the departure intervals per urgency (left); and the long-run average reward improvements over the non-controlled benchmark per urgency process (right).}
\label{fig:max-Nash-welfare-example}
\end{figure}
\begin{mdframed}[backgroundcolor=green!10]
\begin{example}[Allocation of Highway Priority Lanes under MLNW]
\label{ex:running-example-max-Nash-welfare}
Figure~\ref{fig:max-Nash-welfare-example} illustrates the MLNW solution (Figure~\ref{fig:max-Nash-welfare-solution}) for allocating highway priority lanes over the morning rush hour, as specified in Example~\ref{ex:running-example-full}.
We additionally consider that the users are uniformly grouped in one of the four urgency processes shown in Figure~\ref{fig:heterogeneous-urgency-example}, which all have the same average urgency but different dynamics\footnote{Optimization problem~\eqref{eq:max-Nash-welfare} is solved using a representative user of each urgency process, rather than all 9000 users, leveraging the symmetry property of Proposition~\ref{prop:max-Nash-scale-invariance}.}.
To highlight the fairness and efficiency of MLNW, it is contrasted to two other solutions: 
Figure~\ref{fig:static-Nash-welfare-solution} shows the outcomes if Nash welfare is maximized separately for each day in a single-shot manner; and Figure~\ref{fig:utilitarian-solution} shows the utilitarian solution, which assumes that the urgency is interpersonally comparable and allocates the most desirable resources to the highest urgency irrespective of the underlying urgency process.
Contrasting first to the single-shot maximum Nash welfare solution, it allocates all users their \emph{fair share} of each departure interval irrespective of urgency, due to its scale invariance property, cf. Figure~\ref{fig:static-Nash-welfare-solution}--left.
While this still leads to strong Pareto improvements in the long-run expected average rewards over the non-controlled benchmark, cf. Figure~\ref{fig:static-Nash-welfare-solution}--right, it is pronouncedly dominated by MLNW that properly accounts for the dynamics, cf. Figure~\ref{fig:max-Nash-welfare-solution}--right.
It can be seen that MLNW extracts possible efficiency gains from the urgency dynamics in an interpersonally non-comparable, and thereby fairness preserving, manner.
Namely, as revealed by Figure~\ref{fig:max-Nash-welfare-solution}--left, different users `trade' their fair shares of different departure intervals in a mutually beneficial manner.
For example, users who experience urgency $u_\textup{h}=9$ one eighth of the time (displayed green) give up the majority of their fair shares of most departure intervals in exchange of a larger share of the most desirable intervals $j^*$ and $j^*-1$.
On the other hand, users who experience a constant urgency $u=2$ (displayed red), while they give up some of their fair shares of $j^*$, $j^*-1$, $j^*-2$, gain a larger share of all other intervals.

Contrasting second to the utilitarian solution demonstrates the consequences of naively assuming that urgency is interpersonally comparable.
In this case, users who experience a constant urgency $u=2$ (displayed red) are never allocated to the priority lanes, because there are always sufficiently many users with `higher' urgency to fill those lanes, cf. Figure~\ref{fig:utilitarian-solution}--left. As a consequence, those users do not realize any benefits over the non-controlled benchmark, cf. Figure~\ref{fig:utilitarian-solution}--right.
If one interprets the urgency process as how often a user is late to work, those who are always disciplined in adhering to schedules are punished under utilitarianism.
Naturally, in the absence of money or a fair instrument to make interpersonal comparisons, one would expect these users to over-report their urgency to realize benefits under a utilitarian scheme.
\end{example}
\end{mdframed}

%% file: sections/karma.tex
\section{Dynamic Resource Allocation with Karma Economies}
\label{sec:karma}

In this section, we formally introduce \emph{karma economies} for dynamic resource allocation (Section~\ref{sec:karma-description}) and their game-theoretical equilibrium notion of \emph{Karma Equilibrium (KE)} (Section~\ref{sec:karma-macro}).
The main result of this section, which establishes that karma economies implement MLNW in a decentralized and trustworthy manner (Theorem~\ref{thm:Nash-is-KE}), follows in Section~\ref{sec:max-Nash-welfare-karma-equilibrium}.
Moreover, Section~\ref{sec:karma-dpg} discusses a complementary modelling tool for karma economies that produces finer-grained equilibrium predictions, while Section~\ref{sec:related-mechanisms} contrasts karma economies to related mechanisms in the economics and computer science literature, highlighting the `closed-loop' nature of karma.

\subsection{Description of karma economy}
\label{sec:karma-description}
As briefly introduced in Example~\ref{ex:running-example-w-karma} and illustrated in Figure~\ref{fig:running-example-karma}, a karma economy is a mechanism that repeatedly allocates the resources $\M$ to the users $\N$ over infinite time-steps $t \in \Nat$.
Each user $i \in \N$ is endowed with time-varying \emph{karma tokens} $k_{t,i} \in \Real_{\geq 0}$ which it uses to place bids $b_{t,i,j} \in \Real_{\geq 0}$ on the resources $j \in \M$, and the total bid cannot exceed the karma, i.e., $\sum_{j \in \M} b_{t,i,j} \leq k_{t,i}$.
Each resource $j$ is then allocated to the $c_j$-highest bidders, i.e., $x_{t,i,j} = \mathbbm 1\left\{b_{t,i,j} \geq b^\star_{t,j}\right\}$, where $b^\star_{t,j} =\min_{i\in\N}\left\{b_{t,i,j}; \, \lvert i' \in \N; \, b_{t,i',j} > b_{t,i,j} \rvert < c_j \right\}$ denotes the $c_j$-highest bid\footnote{In case of ties on $b^\star_{t,j}$, the remaining capacity after allocating to users bidding strictly greater than $b^\star_{t,j}$ can be allocated randomly.}.
After the allocations are settled, each user $i$ pays the \emph{uniform clearing bid} $b^\star_{t,j}$ of each resource $j$ it receives, i.e., $p_{t,i} = \sum_{j \in \M} x_{t,i,j} \, b^\star_{t,j}$\footnote{Other payment rules are possible, e.g., pay as bid~\citep{elokda2024self,elokda2024carma}; however uniform payments simplify the upcoming exposition.}.
Importantly, at the end of each time-step, the total payment $p^\textup{tot}_{t} = \sum_{i \in \N} p_{t,i} = \sum_{j \in \M} b^\star_{t,j} \, c_j$ is redistributed proportionally to the user-specific access rights $\bm{w}$, leading to a redistribution share for user $i$ of $g_{t,i} = \frac{w_i}{\sum_{i' \in \N} w_{i'}} \: p^\textup{tot}_t$.
Accordingly, each user $i$'s next time-step karma evolves as per $k_{t+1,i} = k_t - p_{t,i} + g_{t,i}$, and the mechanism is repeated indefinitely.
Notice that, due to the redistribution of payments, the average karma in the system is preserved over time, i.e., $\frac{1}{n} \sum_{i \in \N} k_{t,i} = \frac{1}{n} \sum_{i \in \N} k_{t',i} = \bar k$, for all $t, t' \in \N$.
Therefore, to maintain a desired system average $\bar k \in \Real_{>0}$, each user $i$ can be initially endowed with karma $k_{1,i} = \bar k$.



\subsection{Game-theoretical solution concept: Karma Equilibrium (KE)}
\label{sec:karma-macro}

In order to introduce a notion of game-theoretical equilibrium of the karma economy, let us take the point of view of an individual user $i \in \N$ who faces \emph{stationary clearing bids} $\bm{b^\star_t} = \bm{b^\star} \in \Real_{\geq 0}^m$ at all times $t \in \N$.
The stationarity assumption is reasonable in the large population settings envisioned for future socio-technical control.
Namely, assuming that users can be grouped into a finite number of types with identical urgency processes, and that they follow stationary bidding behaviors given their urgency, the realized distribution of bids in every time-step converges to an expected stationary distribution in the large population limit, leading to stationary clearing bids.
Moreover, the distribution of bids becomes dense and an individual user has limited ability to affect the clearing bids in the large population limit.
Under this stationarity assumption, user $i$'s decision problem reduces to whether and at what times it should acquire resources $j \in \M$ by bidding higher than $b^\star_j$ and paying $b^\star_j$, i.e., the user must decide its \emph{long-run allocation} $\bm{\chi_i} = \left(\chi_{i,j}(u_i)\right)_{j \in \M, u_i \in \U_i} \in [0,1]^{m \times q_i}$ specifying the probabilities of accessing the different resources in the different urgency states.
The user is limited in how often it can acquire resources by its karma budget, though.
This is formalized by the following optimization problem.

\begin{mdframed}[backgroundcolor=red!10]
\begin{minipage}{0.16\textwidth}\vspace{0pt}
    \textbf{Karma User Problem:}
\end{minipage}
\hfill
\begin{minipage}{0.83\textwidth}\vspace{0pt}
\begin{equation}
\begin{aligned}
    \max_{\bm{\chi_i}} &&& \bar r_i - \bar r^0_i \\
    \textup{subject to} &&& \sum_{u_i \in \U_i} \sigma_i(u_i) \sum_{j \in \M} \chi_{i,j}(u_i) \, b^\star_j \leq \frac{w_i}{\sum_{i' \in \N} w_{i'}} \sum_{j \in \M} b^\star_j \, c_j, \\
    &&& 0 \leq \chi_{i,j}(u_i) \leq 1, \; \forall j \in \M, u_i \in \U_i.
\end{aligned}
\tag{KU} \label{eq:karma-user-problem}
\end{equation}
\end{minipage}    
\end{mdframed}

In Problem~\eqref{eq:karma-user-problem}, $\bar r_i - \bar r^0_i = \sum_{u_i \in \U_i} \sigma_i[u_i] \, u_i \sum_{j \in \M} \chi_{i,j}(u_i) \left(r_{i,j} - r^0_{i,j}\right)$ (as in Section~\ref{sec:problem-formulation}).
The first constraint captures that the karma budget is satisfied in a \emph{long-run expected average sense}, i.e., user $i$'s expected average karma expenditures are less than or equal to its share of karma redistribution; and the second set of constraints ensure that $\chi_{i,j}(u_i)$ are valid probabilities.
Analogous to Section~\ref{sec:problem-formulation}, in settings where the resources are mutually exclusive, we would additionally require that $\sum_{j \in \M} \chi_{i,j}(u_i) \leq 1$ for all $u_i \in \U_i$.
With this, we are ready to state the definition of the \emph{Karma Equilibrium (KE)}.

\begin{mdframed}[backgroundcolor=red!10]
\begin{definition}[Karma Equilibrium (KE)]
    A Karma Equilibrium (KE) is a pair $\left(\bm{\chi^\star}, \bm{b^\star}\right)$ of long-run allocations $\bm{\chi^\star} = \left(\bm{\chi^\star_i}\right)_{i \in \N}$ and stationary bids $\bm{b^\star} = \left(b^\star_j\right)_{j \in \M}$ that satisfy:
    \begin{enumerate}
        \item \emph{Individual Optimality:} For all users $i \in \N$, $\bm{\chi^\star_i}$ solves Problem~\eqref{eq:karma-user-problem} given $\bm{b^\star}$;

        \item \emph{Resource Clearing:} For all resources $j \in \M$, $\sum\limits_{i \in \N} \sum\limits_{u_i \in \U_i} \sigma_i(u_i) \, \chi^\star_{i,j}(u_i) \leq c_j$, and $b^\star_j > 0$ implies that $\sum\limits_{i \in \N} \sum\limits_{u_i \in \U_i} \sigma_i(u_i) \, \chi^\star_{i,j}(u_i) = c_j$;

        \item \emph{Budget Balance:} For all users $i \in \N$, $\sum\limits_{u_i \in \U_i} \sigma_i(u_i) \sum\limits_{j \in \M} \chi^\star_{i,j}(u_i) \, b^\star_j = \cfrac{w_i}{\sum_{i' \in \N} w_{i'}} \sum\limits_{j \in \M} b^\star_j \, c_j$.
    \end{enumerate}
\end{definition}
\end{mdframed}

A KE thus constitutes the solutions $\bm{\chi^\star}$ of the $n$ karma user problems, which are coupled through stationary bids $\bm{b^\star}$ that correctly clear the resource capacities.
Moreover, all users' expected average expenditures must equal their redistribution shares in the KE, which is needed for stationarity: if for a user $i \in \N$ it holds that $\sum\limits_{u_i \in \U_i} \sigma_i(u_i) \sum\limits_{j \in \M} \chi^\star_{i,j}(u_i) \, b^\star_j < \frac{w_i}{\sum_{i' \in \N} w_{i'}} \sum\limits_{j \in \M} b^\star_j \, c_j$, that user will gather more of the system karma over time leading to constant deflation of $\bm{b^\star}$.

A natural question is if a KE is guaranteed to exist, and in fact it is not.
Namely, it is straightforward to construct cases that violate the budget balance condition.
For example, consider two resources $j \in \{1,2\}$ and a majority of users that desire resource $1$ more than resource $2$, while a small minority desire resource $2$ only.
This will lead to $b^\star_1 > b^\star_2$, and it could occur that $b^\star_2$ is less than a minority user's redistribution share, who is then able to always secure resource $2$ while gaining more karma\footnote{This issue can be addressed by setting a maximum karma cap beyond which users receive no more redistribution, cf. \cite{elokda2024carma}. However, this convolutes the model and analysis.}.
In Section~\ref{sec:max-Nash-welfare-karma-equilibrium}, where we establish the connection of KE to MLNW, we present sufficient conditions for KE to exist.

We write a KE in extended form as $\left(\bm{\chi^\star}, \bm{b^\star},\bm{\eta^\star}\right)$, where additionally $\bm{\eta^\star} = \left(\bm{\eta^\star_i}\right)_{i \in \N}$, $\bm{\eta^\star_i} = \left(\eta^\star_{i,j}(u_i)\right)_{j \in \M, u_i \in \U_i} \in \Real_{\geq 0}^{m \times q_i}$ denote the optimal Lagrange multipliers of the $n$ user problems for constraints $\chi_{i,j}(u_i) \leq 1$.
The long-run expected average reward improvements over the no allocation benchmark at the KE are further denoted by $\left(\bar r_i - \bar r^0_i\right)^\star$.
We have the following characterization of KE.

\begin{mdframed}[backgroundcolor=blue!10]
\begin{proposition}[Properties of KE]
    \label{prop:karma-equilibrium}
    Let Assumption~\ref{as:competition} hold.
    Then KE $(\bm{\chi^\star}, \bm{b^\star}, \bm{\eta^\star})$ satisfy, for all $i \in \N$, $j \in \M$, $u_i \in \U_i$:
    \begin{enumerate}[label=\ref{prop:karma-equilibrium}.\arabic*]
        \item \label{prop:karma-eq-all-improve} $\left(\bar r_i - \bar r^0_i\right)^\star > 0$;

        \item \label{prop:karma-eq-scale-invariance} If two users $i$, $i'$ are symmetric up to scales $\alpha_i, \alpha_{i'} >0$, then $\cfrac{\left(\bar r_i - \bar r^0_i\right)^\star}{\alpha_i} = \cfrac{\left(\bar r_{i'} - \bar r^0_{i'}\right)^\star}{\alpha_{i'}}$;

        \item \label{prop:karma-eq-high-urgency-first} $\chi^\star_{i,j}(u_i) > 0 \Rightarrow \chi^\star_{i,j}(u'_i) = 1$, for all $u'_i > u_i$;
        
        \item \label{prop:karma-eq-resources-used} $b^\star_j > 0$, and all resources are allocated at capacity;
        
        \item \label{prop:karma-eq-stationarity-inequality} $u_i \left(r_{i,j} - r^0_{i,j}\right) \leq \cfrac{\sum_{i'\in \N} w_{i'}}{w_i \sum_{j' \in \M} b^\star_{j'} \, c_{j'}} \left(\left(\bar r_i - \bar r^0_i\right)^\star - \sum\limits_{u'_i \in \U_i} \sum\limits_{j' \in \M} \eta^\star_{i,j'}(u'_i)\right) b^\star_j + \cfrac{\eta^\star_{i,j}(u_i)}{\sigma_i(u_i)}$;

        \item \label{prop:karma-eq-stationarity-equality} $\chi^\star_{i,j}(u_i) > 0 \Rightarrow u_i \left(r_{i,j} - r^0_{i,j}\right)$\\$= \cfrac{\sum_{i'\in \N} w_{i'}}{w_i \sum_{j' \in \M} b^\star_{j'} \, c_{j'}} \left(\left(\bar r_i - \bar r^0_i\right)^\star - \sum\limits_{u'_i \in \U_i} \sum\limits_{j' \in \M} \eta^\star_{i,j'}(u'_i)\right) b^\star_j + \cfrac{\eta^\star_{i,j}(u_i)}{\sigma_i(u_i)}$;

        \item \label{prop:karma-eq-fair-share} If there is only one resource $\M=\{1\}$, then $\sum_{u_i \in \U_i} \sigma(u_i) \chi^\star_{i,1}(u_i) = \frac{w_i}{\sum_{i' \in \N}w_{i'}} \, c_1$.
    \end{enumerate}

    Moreover, if the resources are mutually exclusive, then Propositions~\ref{prop:karma-eq-all-improve}--\ref{prop:karma-eq-scale-invariance} hold, Propositions~\ref{prop:karma-eq-stationarity-inequality}--\ref{prop:karma-eq-stationarity-equality} hold with non-resource specific multipliers $\eta^\star_{i,j}(u_i)=\eta^\star_i(u_i)$, and it holds for all $i \in \N$, $j \in \M$, $u_i \in \U_i$:
    \begin{enumerate}
        \item[\ref{prop:karma-eq-high-urgency-first}']  \label{prop:karma-eq-high-urgency-first-mutually-exclusive} $\chi^\star_{i,j}(u_i) > 0 \Rightarrow \sum_{j' \in \M} \chi^\star_{i,j'}(u'_i) = 1$, for all $u'_i > u_i$;
        
        \item[\ref{prop:karma-eq-resources-used}'] \label{prop:karma-eq-resources-used-mutually-exclusive} $b^\star_j = 0 \Rightarrow \sum_{j' \in \M} \chi^\star_{i,j'}(u_i) = 1$, for all $i \in \N_j$, $u_i \in \U_i$.
    \end{enumerate}
\end{proposition}
\end{mdframed}

The proof of Proposition~\ref{prop:karma-equilibrium} is included in~\ref{app:proof-karma-equilibrium}.
The KE mirrors the properties of MLNW, cf. Proposition~\ref{prop:max-Nash-welfare}, with the equilibrium clearing bids $\bm{b^\star}$ taking a similar role as the shadow prices $\bm{\lambda^*}$ under MLNW.
However, a connection between both solution concepts is not immediate, as comparing Propositions~\ref{prop:karma-eq-stationarity-inequality}--\ref{prop:karma-eq-stationarity-equality} to Propositions~\ref{prop:max-Nash-stationarity-inequality}--\ref{prop:max-Nash-stationarity-equality} reveals: constraints $\chi_{i,j}(u_i) \leq 1$ and the associated optimal multipliers $\eta^\star_{i,j}(u_i)$ affect the solution differently in both concepts.
Moreover, in single resource settings, the KE guarantees that each user's long-run expected average allocation equals its \emph{fair share}, cf. Proposition~\ref{prop:karma-eq-fair-share}, which in general is not guaranteed to hold under MLNW.
In what follows, we establish conditions under which both solution concepts coincide.

\subsection{Trustworthy MLNW implementation with karma economies}
\label{sec:max-Nash-welfare-karma-equilibrium}

\begin{mdframed}[backgroundcolor=red!10]
\begin{definition}[Nash-Balanced Problem]
    \label{def:Nash-is-KE}
    A dynamic resource allocation problem $\DRA$ is said to be \emph{Nash-balanced} if it admits a MLNW solution $\left(\bm{\chi^*},\bm{\lambda^*},\bm{\eta^*}\right)$ satisfying, for all users $i,i' \in \N$
    \begin{align}
        \frac{\sum_{j \in \M} \sum_{u_i \in \U_i} \eta^*_{i,j}(u_i)}{w_i} = \frac{\sum_{j \in \M} \sum_{u_{i'} \in \U_{i'}} \eta^*_{i',j}(u_{i'})}{w_{i'}} = C. \tag{NB} \label{eq:Nash-is-KE}
    \end{align}
\end{definition}
\end{mdframed}

In a Nash-balanced problem, there exists a MLNW solution that penalizes users for satisfying constraints $\chi^*_{i,j}(u_i) \leq 1$ in a manner that is proportional to their access rights.

\begin{mdframed}[backgroundcolor=orange!10]
\begin{theorem}(MLNW is KE)
    \label{thm:Nash-is-KE}
    Let Assumption~\ref{as:competition} hold, and suppose that dynamic resource allocation problem $\DRA$ is Nash-balanced with MLNW solution $(\bm{\chi^*}, \bm{\lambda^*}, \bm{\eta^*})$ satisfying condition~\eqref{eq:Nash-is-KE} and optimal long-run expected average reward improvements $\left(\bar r_i - \bar r^0_i\right)^*_{i \in \N}$.
    Then for all $\alpha > 0$, $\left(\bm{\chi^\star},\bm{b^\star},\bm{\eta^\star}\right)=\left(\bm{\chi^*},\alpha \, \bm{\lambda^*}, \left(\frac{\left(\bar r_i - \bar r^0_i\right)^*}{w_i} \, \bm{\eta^*_i}\right)_{i \in \N}\right)$ is a KE.
\end{theorem}
\end{mdframed}

The proof of Theorem~\ref{thm:Nash-is-KE} is included in~\ref{app:proof-Nash-is-KE}.
The significance of Theorem~\ref{thm:Nash-is-KE} is that it establishes a possibility of attaining MLNW in a game-theoretic equilibrium that does not rely on explicit knowledge of the private urgency processes and reward structures of the users.
Moreover, it provides a proof of existence of a KE for Nash-balanced problems.
Theorem~\ref{thm:Nash-is-KE} can be operationalized in several ways.
First, in an \emph{indirect revelation} implementation, the users participate actively in the karma economy by placing their own bids.
If the underlying problem is Nash-balanced, and if a KE $\left(\bm{\chi^\star},\bm{b^\star},\bm{\eta^\star}\right)$ of Theorem~\ref{thm:Nash-is-KE} is reached, Nash welfare is maximized in a decentralized fashion.
Second, in a \emph{direct revelation} implementation, the users report their private urgency processes and reward structures to a central mechanism in an initial offline step, and subsequently report their private urgency online in each time-step.
The central mechanism computes a KE $\left(\bm{\chi^\star},\bm{b^\star},\bm{\eta^\star}\right)$ of Theorem~\ref{thm:Nash-is-KE} in the offline step, and simulates the karma economy in which it bids according to the KE policy with the online urgency reports.
If the underlying problem is Nash-balanced, and by the \emph{revelation principle}~\cite[Proposition~9.44]{nisan2007introduction}, truthful reporting is (approximately) Bayesian Nash incentive compatible in large population settings in which a single user has limited effect on $\bm{b^\star}$.

It is possible to write a mirror condition of the Nash-balance condition~\eqref{eq:Nash-is-KE}, expressed in terms of the KE multipliers $\bm{\eta^\star}$ rather than the MLNW multipliers $\bm{\eta^*}$, under which the converse of Theorem~\ref{thm:Nash-is-KE} holds, i.e., the KE is guaranteed to maximize long-run Nash welfare.
We present the direction MLNW to KE since it exposes under what classes of urgency process heterogeneity MLNW can be implemented truthfully, as discussed next.

\begin{figure}[!h]
    \centering
    \begin{subfigure}[b]{0.48\textwidth}
        \centering
        \includegraphics[width=0.49\textwidth]{figures/urgency-markov-chain-dynamic.pdf}
        \hfil
        \includegraphics[width=0.49\textwidth]{figures/urgency-markov-chain-very-dynamic.pdf}

        \medskip
        
        \includegraphics{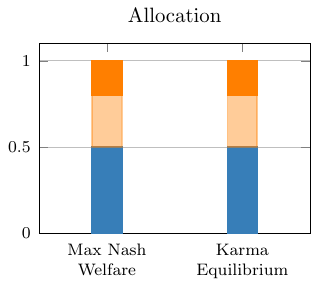}
        \caption{Nash-balanced problem.}
        \label{fig:Nash-is-KE}
    \end{subfigure}
    \hfill
    \begin{subfigure}[b]{0.48\textwidth}
        \centering
        \includegraphics[width=0.49\textwidth]{figures/urgency-markov-chain-dynamic.pdf}
        \hfil
        \includegraphics[width=0.49\textwidth]{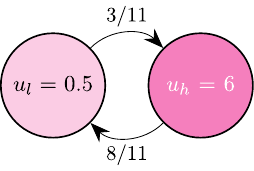}

        \medskip

        \includegraphics{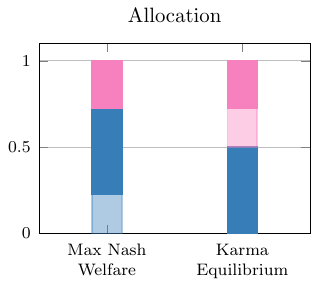}
        \caption{Non-Nash-balanced problem.}
        \label{fig:Nash-not-KE}
    \end{subfigure}

\caption{Comparison of urgency process heterogeneity that leads to Nash-balanced vs. non-Nash balanced dynamic resource allocation problems.}
\label{fig:Nash-is-or-not-KE}
\end{figure}
\begin{mdframed}
\paragraph{\textbf{Discussion of Nash-balance condition}}
While it remains an open problem to fully characterize Nash-balanced dynamic resource allocation problems, we discuss when condition~\eqref{eq:Nash-is-KE} is expected to hold and argue that it is not severely restrictive.
A natural setting in which condition~\eqref{eq:Nash-is-KE} is satisfied is when constraints $\chi^*_{i,j} \leq 1$ do not bind for any $i \in \U_i$, $j \in \M$, $u_i \in \U_i$, and thus $C = \frac{\sum_{u_i \in \U_i} \sum_{j \in \M} \eta^*_{i,j}(u_i)}{w_i}=0$ for all $i \in \N$.
By Proposition~\ref{prop:max-Nash-high-urgency-first}, this occurs when there is resource scarcity leading to MLNW allocations to users in their respective highest urgency only.
Notice that this does not imply that the frequency and severity of the highest urgency must be the same for all users, or that the users end up with equal allocations: in fact, the afore-discussed example problem of allocating highway priority lanes over the morning rush hour (Example~\ref{ex:running-example-max-Nash-welfare}) \emph{is} Nash-balanced, with $\eta^*_{i,j}(u_i)=\eta^*_i(u_i)=0$ for all $i \in \N$, $j \in \M$, $u_i \in \U_i$, and the MLNW solution portrayed in Figure~\ref{fig:max-Nash-welfare-solution} \emph{is indeed a KE of this problem}.

Moreover, there exist Nash-balanced problems for which $C > 0$ and thus constraints $\chi^*_{i,j}(u_i) \leq 1$ bind for each user $i \in \N$ at some resource $j \in \M$ and urgency $u_i \in \U_i$.
Figure~\ref{fig:Nash-is-or-not-KE} shows an example with a single resource $\M=\{1\}$ of capacity $c_1 = 1$, equal access rights, and users belonging to two pairs of urgency processes: one pair leads to a Nash-balanced problem with $C > 0$ (Figure~\ref{fig:Nash-is-KE}) and the other does not lead to a Nash-balanced problem (Figure~\ref{fig:Nash-not-KE}).
By Proposition~\ref{prop:karma-eq-fair-share}, the KE allocates users of both processes the fair share of $0.5$ in both problems, while this occurs under MLNW only in the Nash-balanced problem, cf. Figure~\ref{fig:Nash-is-KE}.
In the non-Nash-balanced problem, it is optimal under MLNW to allocate users of one of the urgency processes more than fair share, cf. Figure~\ref{fig:Nash-not-KE}.
Intuitively, one would not expect this outcome to be implementable in a game-theoretic equilibrium, as users of the urgency process that receives less than fair share would increase their allocation if they over-report the time spent in high urgency.
In mathematical terms, those users are penalized disproportionally for satisfying constraints $\chi^*_{i,j}(u_i) \leq 1$ under MLNW.
\end{mdframed}

\subsection{Model refinement with Dynamic Population Games (DPGs)}
\label{sec:karma-dpg}

The KE $(\bm{\chi^\star}, \bm{b^\star})$ provides a high-level description of the karma economy.
From $(\bm{\chi^\star}, \bm{b^\star})$, one can construct explicit bidding behaviors for each time-step $t \in \Nat$, in which user $i \in \N$ in urgency $u_{t,i} \in \U_i$ bids $b_{t,i,j}=b^\star_j$ (respectively, $b_{t,i,j}=0$) on resources $j \in \M$ with probability $\chi^\star_{i,j}(u_i)$ (respectively, $1 - \chi^\star_{i,j}(u_i)$).
However, since the KE respects the karma budget in a long-run expected average sense, a user may not always have enough karma to realize its bid ex-post.
Moreover, the KE assumes that users are infinitely far-sighted while users may prioritize immediate rewards over future ones.
For this purpose, a complementary \emph{Dynamic Population Game (DPG)} model was developed that gives fine grained insights on the karma economy~\citep{elokda2024self,elokda2024carma}.
DPGs are a tractable subclass of \emph{mean-field games} that enable computing complex and high-dimensional \emph{Stationary Nash Equilibria (SNE)}~\citep{elokda2024dynamic}.
The SNE is a finer-grained solution concept than the KE that fully accounts for the karma budget in every time-step, considers discounted future rewards, and provides detailed bidding policies mapping both urgency \emph{and} karma to (potentially probabilistic) bids.
Moreover, in addition to optimal bidding policies, the SNE predicts the \emph{stationary distribution of karma} in the population, which holds more information than the stationary bids $\bm{b^\star}$.
For example, the SNE allows to quantify the fraction of times users are expected to spend below a certain karma level, as well as whether `karma wealth inequalities' can arise~\citep{elokda2024self}.
The DPG model was applied in multiple extensions, including endogenous redistribution shares that depend on the users' actions, e.g., at what time they depart during the morning rush hour~\citep{elokda2024carma}, elastic population of users~\citep{elokda2023dynamic}, and pairwise or otherwise limited resource contests such as in autonomous driving~\citep{elokda2024self,chavoshi2024introducing}.
However, the DPG model is computational in nature and less amenable to analysis, for which the high-level KE is advantageous.

\subsection{Related mechanisms}
\label{sec:related-mechanisms}

We conclude our exposition of karma economies with a brief discussion of related concepts in economics and computer science that rely on the same principle of \emph{budgeting resource consumption over time}; highlighting the distinguishing feature of karma that it forms a `closed-loop' system that can be sustained indefinitely.
Before discussing mechanisms that budget over time, it is insightful to recall more classical mechanisms that statically \emph{budget over space}.

\begin{mdframed}
\paragraph{\textbf{Budgeting over space: Competitive Equilibrium (CE), Fisher markets, pseudomarkets}}
The Competitive Equilibrium (CE) is a cornerstone of economic theory that models how users trade resources in large markets~\citep{arrow1954existence}.
In full generality, the budget of each user is its \emph{initial endowment of resources}, or the resources it contributes to the market, and the CE is a set of resource prices at which each user is best off selling part or all of its initial endowment and buying resources from others, effectively facilitating efficient trade.
\emph{Fisher markets} are a more closely related special case, in which the resources are public or owned by a central entity and each user is endowed with a budget of artificial currency, representing its access rights to resources~\citep{vazirani2007combinatorial}.
In the case of linear and separately additive utilities, Fisher market CE maximize Nash welfare~\citep[Theorem~5.1]{vazirani2007combinatorial}.
Moreover, when the resources are indivisible, similar schemes go by the name of \emph{pseudomarkets}~\citep{hylland1979efficient} and have been famously applied in the context of allocating course seats to students in business schools~\citep{budish2011combinatorial}.

These mechanisms that budget over space rely on there being a large variety of resources and heterogeneity in needs to attain efficiency gains.
If the users are homogeneous, or if there is a single (most desirable) resource, these mechanisms yield trivial equal splits of resources (if all users have equal access rights).
The time dimension is thus vital to realize efficiency gains beyond those attainable with space alone.
\end{mdframed}

\begin{figure}[!h]
    \centering
    \begin{subfigure}[b]{\textwidth}
        \centering
        \includegraphics[width=\textwidth]{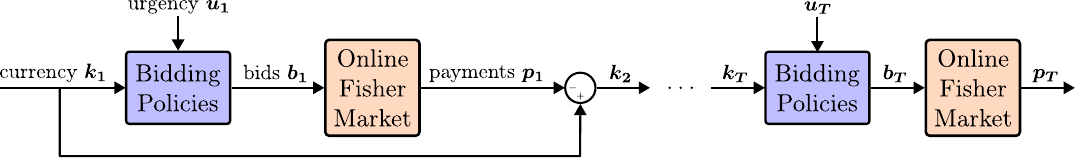}
        \caption{`Open-loop' online Fisher market. Currency is spent over a finite time $T$.}
        \label{fig:artificial-currency-open-loop}
    \end{subfigure}

    \bigskip
    
    \begin{subfigure}[b]{\textwidth}
        \centering
        \includegraphics[width=0.55\textwidth]{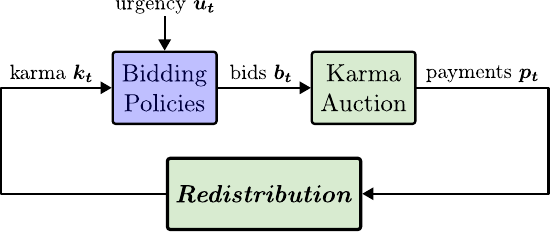}
        \caption{`Closed-loop' karma economy. Karma is spent \emph{and} gained sustainably over infinite time.}
        \label{fig:karma-closed-loop}
    \end{subfigure}
\caption{Comparison of `open-loop' online Fisher markets and `closed-loop' karma economies for dynamic resource allocation.}
\label{fig:artificial-currency-vs-karma}
\end{figure}
\begin{mdframed}
\paragraph{\textbf{Budgeting over time: Linking decisions, online Fisher markets}}
One may argue that time is not fundamentally different from space: if a resource is repeatedly allocated over $T$ time-steps, this can be `converted' to space by allocating $T$-copies of the resource in one shot.
But the time dimension deserves a separate treatment for at least two reasons.
First is the \emph{online} nature of time: users must decide their allocation one time-step at a time, with imperfect knowledge of their future preferences.
Second, the analogy breaks down if the resources are \emph{allocated indefinitely}, i.e., $T \rightarrow \infty$, as this would necessitate infinite initial endowments in the classical spacial models.

Works on budgeting over time draw their roots in the linking mechanism by \cite{jackson2007overcoming} discussed in Section~\ref{sec:pillar-dynamics}.
The linking mechanism was extended to Markovian rather than i.i.d. preference dynamics in \cite{escobar2013efficiency}, which also assumes public knowledge of the dynamic processes and tailors the mechanism to these processes over finite time periods.
The most closely related mechanisms that do not require public knowledge of the preference dynamics are those based on \emph{artificial currency}, also known as \emph{online Fisher markets} as they form the natural extension of static Fisher markets~\citep{gorokh2021remarkable,gao2021online,siddartha2023robust,jalota2025stochastic,lin2025robust}.
In these mechanisms, each user receives an initial endowment of artificial currency to spend over a finite horizon.
While research on online Fisher markets remains active, we remark that they share the same shortcoming of linking mechanisms of being `open-loop.'
The lack of feedback, as facilitated in karma economies by the \emph{redistribution of karma}, makes them not well-suited for infinite resource repetitions; and sensitive to transient shocks or bidding mistakes that could prematurely deplete the budget with no opportunity for recovery.
Figure~\ref{fig:artificial-currency-vs-karma} illustrates this distinguishing feature of karma.
\end{mdframed}

\begin{mdframed}
\paragraph{\textbf{Budgeting over time and space: Feeding America's food allocation system}}
Finally, one example of a mechanism that budgets over time \emph{and} space is Feeding America's \emph{choice system} for allocating food donations to food banks in the United States~\citep{prendergast2022allocation}.
The choice system is essentially a karma economy that was implemented in the real-world: karma tokens are called `shares'; and food banks bid shares for available food donations, pay their bids upon winning, and the total payment is redistributed at the end of each day in proportion to exogenous access rights representing the number of serviced clients per food bank.
This karma-like economy has been widely celebrated for successfully tailoring the allocations to the food banks' needs, leading to an unprecedented fluidity of donations; as well as gaining the trust of all stakeholders involved who perceived the system as fair.
However, the choice system is modeled as a static CE that is repeated daily, under the assumption that food banks would spend all their shares each day; thereby failing to capture the important time dimension~\citep{prendergast2022allocation}.
Indeed, the empirical data shows that food banks spend a small portion of their shares every day to save shares for times of high urgency as characterized by the scarcity of external donations.
\end{mdframed}

%% file: sections/outlook.tex
\section{Outlook: To Couple or Not to Couple? A Smart City Run on Multi-Karma Economies}
\label{sec:outlook-multi-karma}

\begin{figure}[!tb]
    \centering
    \begin{subfigure}[b]{\textwidth}
        \centering
        \includegraphics[width=0.75\textwidth]{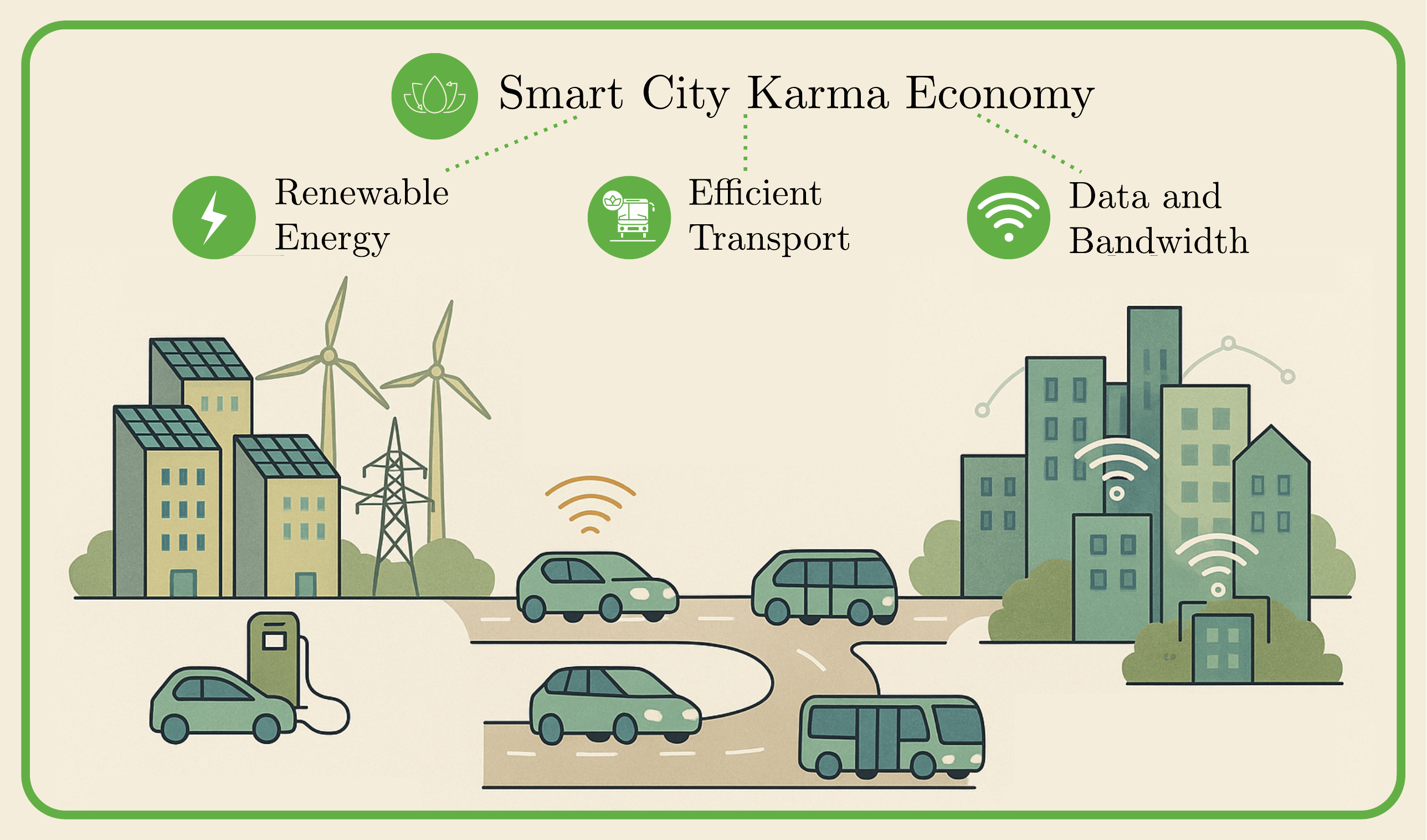}
        \caption{Single global karma economy.}
        \label{fig:smart-city-karma-global}
    \end{subfigure}

    \bigskip
    
    \begin{subfigure}[b]{\textwidth}
        \centering
        \includegraphics[width=0.75\textwidth]{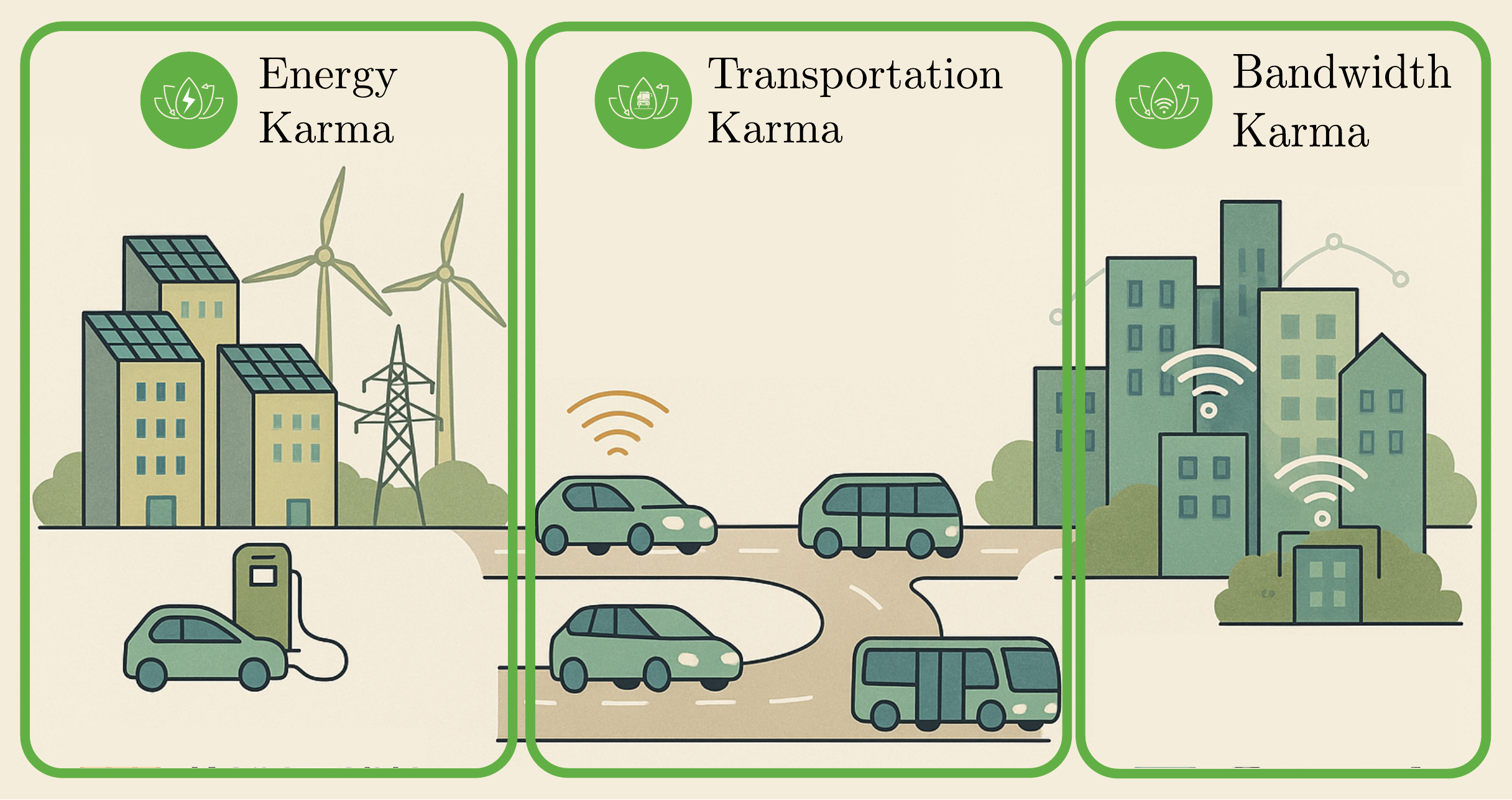}
        \caption{Multiple local karma economies.}
        \label{fig:smart-city-karma-local}
    \end{subfigure}

\caption{Karma economies allow to control the \emph{scope of fairness and efficiency}: Should the future smart city be \emph{globally fair and efficient} across all resource domains or \emph{locally fair and efficient} in different domains?}
\label{fig:smart-city-karma}
\end{figure}

Thus far, we made the case for long-run Nash welfare to jointly formalize \emph{fairness and efficiency} in socio-technical control problems, and showed that it can be maximized in a \emph{trustworthy} manner with karma economies.
An important question remains to settle: \emph{at what scope should fairness and efficiency, i.e., long-run Nash welfare, be maximized?}
Figure~\ref{fig:smart-city-karma} illustrates two possible futures of the smart city.
In the first future city, cf. Figure~\ref{fig:smart-city-karma-global}, long-run Nash welfare is maximized in a \emph{global} sense, and correspondingly, there is a single karma economy for all socio-technical resource allocation needs.
This means that users have full freedom to use the karma gained in one resource domain for another.
For example, a user that participates in electricity demand response and shifts its load off peak times can use its gained electricity karma to access highway priority lanes more frequently.
Similarly, a user that gives up its fair share of bandwidth and settles for slower internet can gain its right to more flexibility on when to charge its Electric Vehicle (EV).
On the other hand, in the second future city, cf. Figure~\ref{fig:smart-city-karma-local}, long-run Nash welfare is maximized \emph{locally} across different domains, and correspondingly, there are multiple separated karma economies per domain.
For example, there is a \emph{transportation karma economy} combining priority roads, parking spaces, micro-mobility, etc., and an \emph{energy karma economy} covering electricity and heating needs, etc..

Conventional wisdom would argue for \emph{free trade}~\citep{friedman2017price} and in favor of the global karma economy of Figure~\ref{fig:smart-city-karma-global}.
For example, it has been recently noted the EVs could provide unique capabilities in future power grids as they are essentially \emph{portable energy storage systems}~\citep{he2021utility}, and recent works have studied coupled incentives for EV charging and traffic routing~\citep{alizadeh2016optimal,xu2018planning,cenedese2022incentive,vcivcic2025active}.
More generally, it is possible to show that users unanimously benefit from combining different karma economies, in theory.
We present this result here for the simplified case of two resources.

\begin{mdframed}[backgroundcolor=blue!10]
\begin{proposition}(Coupling Karma Economies Benefits All)
\label{prop:coupling}
    Suppose that there are two resources $\M={1,2}$, and let $\bar r_i^{\star,1}$ (respectively, $\bar r_i^{\star,2}$) denote user $i$'s long-run expected average reward derived from resource $1$ (respectively, resource $2$) at the KE of two separate karma economies.
    If a KE of the combined karma economy exists, then it satisfies, for all $i \in \N$
    \begin{align*}
        \bar r_i^\star \geq \bar r_i^{\star,1} + \bar r_i^{\star,2},
    \end{align*}
    where $\bar r_i^\star$ denotes user $i$'s long-run expected average reward derived from both resources in the combined economy.
\end{proposition}
\end{mdframed}

The proof of Proposition~\ref{prop:coupling} is included in~\ref{app:proof-coupling}.
It is a direct consequence of Proposition~\ref{prop:karma-eq-fair-share}: at the KE of the single resource economies, each user receives its fair share of each resource on average; and this is guaranteed to remain feasible when combining the economies.
Consequently, each user can only improve by giving up some of its fair share of the less desirable resource for a larger share of the more desirable one.

In light of the theoretical advantages of combining karma economies, is there a case for keeping some economies separate ala Figure~\ref{fig:smart-city-karma-local}?
A lesson learned from classical monetary economies is that despite the theoretical advantages of free trade, many modern democracies have felt a strong need to regulate some markets~\citep{antweiler2001free,irwin2020free}.
The typical concerns are that free trade can indeed make some persons worse off if there are unmodelled externalities, and it can moreover exacerbate inequities.
Similarly, in the future smart city, there could be a strong societal sentiment to keep some resource domains separated from each other.
Perhaps one resource is deemed as more \emph{essential}, and while there could be a short-term benefit for a user to give up its fair share of that resource for another, there could also be an unmodelled long-term harm.
For example, if one couples energy for heating with transportation, it might push some marginalized communities to accept colder living conditions in exchange for faster daily travel, leading to unmodelled and potentially detrimental long-term health consequences.

Moreover, the \emph{time dimension} leveraged by karma economies provides flexibility to \emph{rely less on space}.
As elaborated in Section~\ref{sec:related-mechanisms}, in classical single-shot and static models, the only possibility for efficiency gains is to couple resources through space, i.e., jointly allocate a large variety of resources to leverage heterogeneity in preferences.
With karma economies, while coupling through space also leads to theoretical efficiency gains, cf. Proposition~\ref{prop:coupling}, there are meaningful and pronounced gains to be made with time alone.
This could lead to less severe trade-offs faced by policy-makers in cases where the benefits of combining resource domains are contested.

Finally, in addition to \emph{if} different karma economies should be coupled there is a question of \emph{how} to implement such coupling.
Inspired by monetary economies, one could envision introducing \emph{exchange rates} between different economies, e.g., one unit of transportation karma exchanges for two units of electricity karma; however, it has been shown that exchange rates are inconsequential in equilibrium as they are counter-acted by the scale of bids~\citep{elokda2024travel}.
Moreover, if the goal is to maximize the combined long-run Nash welfare, then Theorem~\ref{thm:Nash-is-KE} dictates the straightforward coupling in which there is simply one karma account to use for all resources.
Another open problem is if users have \emph{different access rights} to the different resources, e.g., representing different initial investments or contributions in maintenance efforts.
To the extent of our knowledge, Nash welfare with intrapersonally different weights on different components of the reward has not been formalized before, and correspondingly, it is not clear how to tailor a coupled karma economy to such a setting.



%% file: sections/conclusion.tex
\section{Conclusion}
\label{sec:conclusion}

To conclude, in this paper, we showed that it is possible to attain maximum long-run Nash welfare, which jointly embodies \emph{fairness and efficiency} in infinitely repeated and dynamic resource allocation settings, in a truth-revealing and thereby \emph{trustworthy} manner, using karma economies.
This stands in contrast to the static impossibility of attaining Pareto efficient resource allocations in a trustworthy manner, without money, in single-shot and static settings.
With this innovation, socio-technical control solutions are equipped to deliver their promise for automated and sustainable future smart cities, in a manner that is better aligned with the human and societal factors of trust, fairness, and efficiency.

Future research should prioritize bringing our vision closer to reality, for which we identify the following directions as particularly important, in addition to the design of multi-karma economies discussed in Section~\ref{sec:outlook-multi-karma}.

\paragraph{Applications}
As a novel general resource allocation method, and as pointed out in Section~\ref{sec:outlook-multi-karma}, karma economies could impact many socio-technical application domains, including the canonical examples of transportation, energy, bandwidth, data, etc..
It is particularly important to study how karma can contribute to climate actions and the pressing clean energy transition.
There is a plethora of problems to explore, ranging from karma-based EV smart charging / peak-shaving; resource sharing in energy communities; fair ways to elicit demand response and provide flexibility for renewable generation; all the way to a completely re-imagined, karma-based future energy market with 100\% renewable and publicly available clean energy.

\paragraph{User interfaces and automated karma bidders}
Given the frequency and complexity of karma bidding, it would be unreasonable to expect citizens of the future smart city to participate in karma economies directly.
The design of simple user interfaces, as well as automated bidders that bid on the users' behalf in real-time, are therefore important directions for future research that will enable scaling the present approach to more complex real-world settings.
In this direction, preliminary research has subjected karma economies to the behavioral test with real humans~\citep{elokda2024behavioral}, as well as investigated adopting automated bidders for monetary ad auctions to the non-monetary karma setting~\citep{berriaud2024spend}.
The active research area of \emph{multi-agent reinforcement learning}~\citep{zhang2021multi} is also relevant for this challenge.

\paragraph{Non-stationary environments}
A standing assumption of this paper is that resource capacities and reward structures are time-invariant, which allows studying the compact equilibrium concept of Karma Equilibrium (KE) expressed in terms of stationary long-run allocations and clearing bids.
An important future extension thus regards relaxing this assumption and allowing resource capacities and reward structures to vary over time, e.g., due to weather conditions, seasonal changes, or disturbance events.
A promising avenue in this direction, that was explored in~\cite{elokda2024travel}, is to introduce a global environment state characterizing changing conditions, and define KE in terms of stationary long-run allocations and clearing bids \emph{conditioned} on the environment state.

\paragraph{Benchmark allocation for long-run Nash welfare}
Our central social welfare measure of long-run Nash welfare relies on a benchmark or status quo allocation $\bm{\chi^0}$ in order to be well-posed in the absence of interpersonal comparability, cf. Section~\ref{sec:Nash-welfare}.
In this paper, we adopted \emph{no allocation} as the benchmark, while it will be interesting to investigate the sensitivity of MLNW allocations to different benchmarks in the future.
The natural choice of benchmark is likely domain-specific, and moreover, the benchmark could have \emph{external inequities} that will persist in the MLNW allocation if not explicitly accounted for (e.g., using the access rights $\bm{w}$).
In the example of allocating highway priority lanes, some marginalized users could have higher travel delays in the benchmark because they have to travel longer distances; and it is an open policy question whether to explicitly prioritize such users in order to correct for this external inequity.
Notice, however, that making claims about the equity of the benchmark requires interpersonal comparisons.

\paragraph{Elastic urgency processes and connection to Nash-balance}
Finally, throughout this paper, we made a standing assumption that the users' urgency processes are fixed and exogenous.
However, users might adjust their urgency processes to the control scheme in an elastic manner, representing, e.g., how often they can afford to be late.
This requires studying \emph{meta-games} in which users also choose their urgency processes.
We conjecture that only urgency processes that make the dynamic resource allocation problem Nash-balanced can be sustained in the equilibria of these meta-games.

%% file: sections/proofs.tex
\section{Proofs}
\label{app:proofs}

\subsection{Proof of Proposition~\ref{prop:max-Nash-welfare}}
\label{app:proof-max-Nash-welfare}

We first derive the KKT conditions of Problem~\eqref{eq:max-Nash-welfare}. The Lagrangian is given by
\begin{align*}
    L(\bm \chi, \bm \lambda, \bm \eta, \bm \iota) =& -\sum_{i \in \N} w_i \log \left(\sum_{u_i \in \U_i} \sigma_i(u_i) \, u_i \sum_{j \in \M} \chi_{i,j}(u_i) \, \left(r_{i,j} -r^0_{i,j}\right) \right) \\
    &\quad + \sum_{j \in \M} \lambda_j \left(\sum_{i \in \N} \sum_{u_i \in \U_i} \sigma_i(u_i) \, \chi_{i,j}(u_i) - c_j \right) \\
    &\quad + \sum_{i \in \N} \sum_{u_i \in \U_i} \sum_{j \in \M} \eta_{i,j}(u_i) \left(\chi_{i,j}(u_i) - 1\right) \\
    &\quad - \sum_{i \in \N} \sum_{u_i \in \U_i} \sum_{j \in \M} \iota_{i,j}(u_i) \, \chi_{i,j}(u_i),
\end{align*}
which leads to the following KKT conditions, for all $i \in \N$, $j \in \M$, $u_i \in \U_i$,
\begin{align*}
    \textup{\textbf{Stationarity:}} \quad & -w_i \, \frac{\sigma_i(u_i) \, u_i \, (r_{i,j} - r^0_{i,j})}{\left(\bar r_i - \bar r^0_i\right)^*} + \sigma_i(u_i) \, \lambda^*_j + \eta^*_{i,j}(u_i) - \iota^*_{i,j}(u_i) = 0; \\[2mm]
    \textup{\textbf{Primal feas.:}} \quad &\sum_{i \in \N} \sum_{u_i \in \U_i} \sigma_i(u_i) \, \chi^*_{i,j}(u_i) \leq c_j, \quad 0 \leq \chi^*_{i,j}(u_i) \leq 1; \\[2mm]
    \textup{\textbf{Dual feas.:}} \quad &\lambda^*_j \geq 0, \quad \eta^*_{i,j}(u_i) \geq 0, \quad \iota^*_{i,j}(u_i) \geq 0; \\[2mm]
    \textup{\textbf{Comp. slack.:}} \quad &\lambda^*_j \left(\sum_{i \in \N} \sum_{u_i \in \U_i} \sigma_i(u_i) \, \chi^*_{i,j}(u_i) - c_j\right) = 0, \\
    \quad & \eta^*_{i,j}(u_i) \left(\chi^*_{i,j}(u_i) - 1\right) = 0, \quad \iota^*_{i,j}(u_i) \, \chi^*_{i,j}(u_i) = 0.
\end{align*}


To show Proposition~\ref{prop:max-Nash-all-improve},
it suffices to show that there is a feasible long-run allocation at which $\left(\bar r_i - \bar r^0_i \right) > 0$ for all $i \in \N$.
Since $\left(\bar r_i - \bar r^0_i \right)$ is bounded for all $i \in \N$, and $\lim_{x \rightarrow 0^+} \log x = -\infty$, such an allocation would attain a higher objective than any allocation in which $\left(\bar r_i - \bar r^0_i \right) \leq 0$ for some $i \in \N$.
Under Assumption~\ref{as:competition}, one such allocation is $\chi_{i,j}(u_i) = \frac{c_j}{\left\lvert\N_j\right\rvert}$ for all $j \in \M$, $i \in \N_j$, $u_i \in \U_i$.
Moreover, uniqueness follows from the strict concavity of the logarithmic objective.

We show Proposition~\ref{prop:max-Nash-scale-invariance} in two steps.
First, if $i$ and $i'$ are symmetric with the same scale, i.e., $\alpha_i$ = $\alpha_{i'}$, it must hold that $\left(\bar r_i - \bar r^0_i\right)^* = \left(\bar r_{i'} - \bar r^0_{i'}\right)^*$ by the uniqueness of $\left(\bar r_i - \bar r^0_i\right)^*$; otherwise a different optimal $\left(\bar r_i - \bar r^0_i\right)^*$ is attained by swapping the allocations of $i$ and $i'$.
Second, the logarithmic objective is scale-invariant, i.e., 
\begin{multline*}
    w_i \left(\log\left(\bar r_i - \bar r^0_i\right) + \log\left(\bar r_{i'} - \bar r^0_{i'}\right)\right) = w_i \left(\log \alpha_i \frac{\left(\bar r_i - \bar r^0_i\right)}{\alpha_i} + \log \alpha_{i'} \frac{\left(\bar r_{i'} - \bar r^0_{i'}\right)}{\alpha_{i'}}\right) \\
    = w_i \left(\log \frac{\left(\bar r_i - \bar r^0_i\right)}{\alpha_i} + \log \frac{\left(\bar r_{i'} - \bar r^0_{i'}\right)}{\alpha_{i'}} + \log \alpha_i + \log \alpha_{i'}\right),
\end{multline*}
and the affine offset $w_i \left(\log \alpha_i + \log \alpha_{i'}\right)$ does not affect the optimum.
Since both $i$ and $i'$ are symmetric with the same scale once normalized by $\alpha_i$ and $\alpha_{i'}$, respectively, it must hold that $\frac{\left(\bar r_i - \bar r^0_i\right)^*}{\alpha_i} = \frac{\left(\bar r_{i'} - \bar r^0_{i'}\right)^*}{\alpha_i'}$.

Propositions~\ref{prop:max-Nash-stationarity-inequality}--\ref{prop:max-Nash-stationarity-equality} follow from straightforward manipulation of the KKT conditions.

To show Proposition~\ref{prop:max-Nash-high-urgency-first}, suppose, for the sake of contradiction, that there exists $i \in \N$, $j \in \M$, $u_i \in \U_i$ for which $\chi^*_{i,j}(u_i) > 0$, and simultaneously $u'_i > u_i$ for which $\eta^*_{i,j}(u'_i)=0$.
Then, Propositions~\ref{prop:max-Nash-stationarity-inequality}--\ref{prop:max-Nash-stationarity-equality} yield
\begin{align*}
    u_i &= \frac{\left(\bar r_i - \bar r^0_i\right)^*}{w_i \left(r_{i,j} - r^0_{i,j}\right)} \left(\lambda^*_j + \frac{\eta^*_{i,j}(u_i)}{\sigma_i(u_i)} \right) \geq \frac{\left(\bar r_i - \bar r^0_i\right)^*}{w_i \left(r_{i,j} - r^0_{i,j}\right)} \, \lambda^*_j \geq u'_i,
\end{align*}
which contradicts $u'_i > u_i$. Therefore it must hold that $\eta^*_{i,j}(u'_i) > 0$, and, by complementary slackness, $\chi^*_{i,j}(u'_i) = 1$.

To show Proposition~\ref{prop:max-Nash-resources-used}, suppose, for the sake of contradiction, that there exists $j \in \M$ for which $\lambda^*_j = 0$.
By Proposition~\ref{prop:max-Nash-stationarity-inequality}, it must hold for all $i \in \N_j$, $u_i \in \U_i$,
\begin{align*}
    \eta^*_{i,j}(u_i) \geq \frac{w_i \, \sigma_i(u_i) \, u_i \left(r_{i,j} - r^0_{i,j}\right)}{\left(\bar r_i - \bar r^0_i\right)^*} > 0,
\end{align*}
and, by complementary slackness, $\chi^*_{i,j}(u_i) = 1$.
But this violates resource $j$'s capacity constraint, by Assumption~\ref{as:competition-resource}, and leads to a contradiction.
The fact that all resources are allocated at capacity then follows from complementary slackness.

Finally, multiplying the stationarity condition by $\chi^*_{i,j}(u_i)$, and using complementary slackness, yields for all $i \in \N$, $j \in \M$, $u_i \in \U_i$
\begin{align*}
    \sigma_i(u_i) \, \chi^*_{i,j}(u_i) \, \lambda^*_j + \eta^*_{i,j}(u_i) = w_i \, \frac{\sigma_i(u_i) \, u_i \, \chi^*_{i,j}(u_i) \left(r_{i,j} - r^0_{i,j}\right)}{\left(\bar r_i - \bar r^0_i\right)^*},
\end{align*}
and summing over $j \in \M$, $u_i \in \U_i$ yields Proposition~\ref{prop:max-Nash-weight}.

The proof for the mutually exclusive resource setting follows analogous arguments.

\qed

\subsection{Proof of Proposition~\ref{prop:karma-equilibrium}}
\label{app:proof-karma-equilibrium}

We first derive the $n$ KKT systems that are necessary and sufficient for the optimality of Problems~\eqref{eq:karma-user-problem}.
For user $i \in \N$, the Lagrangian is given by
\begin{align*}
    L_i(\bm{\chi_i}, \kappa_i,\bm{\eta_i},\bm{\iota_i}) =& -\sum_{u_i \in \U_i} \sigma_i(u_i) \, u_i \sum_{j \in \M} \chi_{i,j}(u_i) \, \left(r_{i,j} -r^0_{i,j}\right) \\
    &\quad + \kappa_i \left(\sum_{u_i \in \U_i} \sigma_i(u_i) \sum_{j \in \M} b^\star_j \, \chi_{i,j}(u_i) - \frac{w_i}{\sum_{i' \in \N} w_{i'}} \sum_{j \in \M} b^\star_j \, c_j \right) \\
    &\quad + \sum_{u_i \in \U_i} \sum_{j \in \M} \eta_{i,j}(u_i) \left(\chi_{i,j}(u_i) - 1\right) \\
    &\quad - \sum_{u_i \in \U_i} \sum_{j \in \M} \iota_{i,j}(u_i) \, \chi_{i,j}(u_i),
\end{align*}
which leads to the following KKT conditions, for all $i \in \N$, $j \in \M$, $u_i \in \U_i$,
\begin{align*}
    \textup{\textbf{Stationarity:}} \quad & -\sigma_i(u_i) \, u_i \, (r_{i,j} - r^0_{i,j}) + \sigma_i(u_i) \, b^\star_j \, \kappa^\star_i + \eta^\star_{i,j}(u_i) - \iota^\star_{i,j}(u_i) = 0; \\[2mm]
    \textup{\textbf{Primal feas.:}} \quad &\sum_{u_i \in \U_i} \sigma_i(u_i) \sum_{j \in \M} b^\star_j \, \chi^\star_{i,j}(u_i) \leq \frac{w_i}{\sum_{i' \in \N} w_{i'}} \sum_{j \in \M} b^\star_j \, c_j, \\
    & 0 \leq \chi^\star_{i,j}(u_i) \leq 1; \\[2mm]
    \textup{\textbf{Dual feas.:}} \quad &\kappa^\star_i \geq 0, \; \eta^\star_{i,j}(u_i) \geq 0, \; \iota^\star_{i,j}(u_i) \geq 0; \\[2mm]
    \textup{\textbf{Comp. slack.:}} \quad &\kappa^\star_i \left(\sum_{u_i \in \U_i} \sigma_i(u_i) \sum_{j \in \M} \chi^\star_{i,j}(u_i) \, b^\star_j  - \frac{w_i}{\sum_{i' \in \N} w_{i'}} \sum_{j \in \M} b^\star_j \, c_j \right) = 0, \\
    & \eta^\star_{i,j}(u_i) \left(\chi^\star_{i,j}(u_i) - 1\right) = 0, \; \iota^\star_{i,j}(u_i) \, \chi^\star_{i,j}(u_i) = 0.
\end{align*}


To show Proposition~\ref{prop:karma-eq-all-improve}, it suffices to show that for each user $i \in \N$, there is a feasible long-run allocation at which $\left(\bar r_i - \bar r^0_i\right) > 0$.
Under Assumption~\ref{as:competition-agent}, it is straightforward to construct such an allocation, e.g., $\chi_{i,j}(u_i) = \epsilon$ for all $j \in \M$ for which $i \in \N_j$, $u_i \in \U_i$, and some sufficiently small $\epsilon > 0$ for which the karma budget constraint is satisfied.

Proposition~\ref{prop:karma-eq-scale-invariance} follows from the fact that any two symmetric users $i, i'$ up to scales $\alpha_i, \alpha_{i'}$ face exactly the same optimization problem with different scaling of the objective.
Therefore, if without loss of generality $\frac{\left(\bar r_i - \bar r_i^0\right)^\star}{\alpha_i} > \frac{\left(\bar r_{i'} - \bar r_{i'}^0\right)^\star}{\alpha_i'}$, user $i'$ long-run allocation cannot be optimal, leading to a contradiction of the KE individual optimality condition.

We next show Propositions~\ref{prop:karma-eq-stationarity-inequality}--\ref{prop:karma-eq-stationarity-equality}.
Multiplying the stationarity conditions by $\chi^\star_{i,j}(u_i)$, and using complementary slackness, yields
\begin{align*}
    \sigma_i(u_i) \, u_i \, \chi^\star_{i,j}(u_i) \left(r_{i,j} - r^0_{i,j}\right) = \kappa^\star_i \, \sigma_i(u_i) \,  \chi^\star_{i,j}(u_i) \, b^\star_j + \eta^\star_{i,j}(u_i).
\end{align*}
Summing over $u_i \in \U_i$, $j \in \M$, and using complementary slackness, yields
\begin{align*}
    &\left(\bar r_i - \bar r^0_i\right)^\star = \kappa^\star_i \, \frac{w_i \sum_{j \in \M} b^\star_j \, c_j}{\sum_{i'\in \N} w_{i'}} + \sum_{u_i \in \ \U_i} \sum_{j \in \M} \eta^\star_{i,j}(u_i) \\
    \Leftrightarrow \quad & \kappa^\star_i = \frac{\sum_{i'\in \N} w_{i'}}{w_i \sum_{j \in \M} b^\star_j \, c_j} \left(\left(\bar r_i - \bar r^0_i\right)^\star - \sum_{u_i \in \U_i} \sum_{j \in \M} \eta^\star_{i,j}(u_i) \right).
\end{align*}
Propositions~\ref{prop:karma-eq-stationarity-inequality}--\ref{prop:karma-eq-stationarity-equality} then follow straightforwardly from stationarity, dual feasibility and complementary slackness.

To show Proposition~\ref{prop:karma-eq-high-urgency-first}, suppose, for the sake of contradiction, that there exists $i \in \N$, $j \in \M$, $u_i \in \U_i$ for which $\chi^\star_{i,j}(u_i) > 0$, and simultaneously $u'_i > u_i$ for which $\eta^\star_{i,j}(u'_i)=0$.
Then, Propositions~\ref{prop:karma-eq-stationarity-inequality}--\ref{prop:karma-eq-stationarity-equality} yield
\begin{multline*}
    u_i = \cfrac{\sum_{i'\in \N} w_{i'}}{w_i \left(r_{i,j} - r^0_{i,j}\right) \sum_{j' \in \M} b^\star_{j'} \, c_{j'}} \left(\left(\bar r_i - \bar r^0_i\right)^\star - \sum\limits_{u''_i \in \U_i} \sum\limits_{j' \in \M} \eta^\star_{i,j'}(u''_i)\right) b^\star_j \\
    + \cfrac{\eta^\star_{i,j}(u_i)}{\sigma_i(u_i) \left(r_{i,j} - r^0_{i,j}\right)} \\
    \geq \cfrac{\sum_{i'\in \N} w_{i'}}{w_i \left(r_{i,j} - r^0_{i,j}\right) \sum_{j' \in \M} b^\star_{j'} \, c_{j'}} \left(\left(\bar r_i - \bar r^0_i\right)^\star - \sum\limits_{u''_i \in \U_i} \sum\limits_{j' \in \M} \eta^\star_{i,j'}(u'_i)\right) b^\star_j \geq u'_i,
\end{multline*}
which contradicts $u'_i > u_i$. Therefore it must hold that $\eta^\star_{i,j}(u'_i) > 0$, and, by complementary slackness, $\chi^\star_{i,j}(u'_i) = 1$.

To show Proposition~\ref{prop:karma-eq-resources-used}, suppose, for the sake of contradiction, that there exists $j \in \M$ for which $b^\star_j = 0$.
By Proposition~\ref{prop:karma-eq-stationarity-inequality}, it must hold for all $i \in \N_j$, $u_i \in \U_i$,
\begin{align*}
    \eta^\star_{i,j}(u_i) \geq \sigma_i(u_i)  \, u_i \left(r_{i,j} - r^0_{i,j}\right) > 0,
\end{align*}
and, by complementary slackness, $\chi^\star_{i,j}(u_i) = 1$.
But this violates resource $j$'s capacity constraint, by Assumption~\ref{as:competition-resource}, and leads to a contradiction of the KE resource clearing condition.
The same condition thus implies that all resources are allocated at capacity.

Finally, Proposition~\ref{prop:karma-eq-fair-share} is immediate from the budget balance condition of KE when there is a single resource $\M=\{1\}$.

The proof for the mutually exclusive resource setting follows analogous arguments.

\qed

\subsection{Proof of Theorem~\ref{thm:Nash-is-KE}}
\label{app:proof-Nash-is-KE}

First, the \emph{resource clearing} condition of KE is satisfied immediately by the fact that $\left(\bm{\chi^*},\bm{\lambda^*},\bm{\eta^*}\right)$ solves Problem~\eqref{eq:max-Nash-welfare}, and by complementary slackness of $\bm{\lambda^*}$.

Second, we verify the \emph{budget balance} condition of KE.
Proposition~\ref{prop:max-Nash-weight} and the Nash-balance condition~\eqref{eq:Nash-is-KE} yield for all users $i \in \N$
\begin{align*}
    \sum_{u_i \in \U_i} \sigma_i(u_i) \sum_{j \in \M} \chi^*_{i,j}(u_i) \, \lambda^*_j = w_i - \sum_{j \in \M} \sum_{u_i \in \U_i} \eta^*_{i,j}(u_i) = (1 - C) \, w_i.
\end{align*}
Summing over all users, and using complementary slackness, yields
\begin{align*}
    &\sum_{j \in \M} \sum_{i \in \N} \sum_{u_i \in \U_i} \sigma_i(u_i) \chi^*_{i,j}(u_i) \, \lambda^*_j = \sum_{j \in \M} \lambda^*_j \, c_j = (1 - C) \sum_{i \in \U_i} w_i. \\
    \Leftrightarrow \quad & 1 - C = \frac{\sum_{j \in \M} \lambda^*_j \, c_j}{\sum_{i \in \U_i} w_i}.
\end{align*}
It follows for all users $i \in \N$
\begin{align*}
    \sum_{u_i \in \U_i} \sigma_i(u_i) \sum_{j \in \M} \chi^*_{i,j}(u_i) \, b^\star_j = \alpha \, \frac{w_i}{\sum_{i'\in\N}w_{i'}} \sum_{j \in \M} \lambda^*_j \, c_j = \frac{w_i}{\sum_{i'\in\N}w_{i'}} \sum_{j \in \M} b^\star_j \, c_j,
\end{align*}
as desired.

Finally, we show that $\bm{\chi^\star_i}$ is optimal given $\bm{b^\star}$ for all users $i\in\N$, by verifying the necessary and sufficient KKT conditions of the $n$ Problems~\eqref{eq:karma-user-problem}.
It is straightforward to verify primal feasibility, dual feasibility, and complimentary slackness; we thus focus on showing that stationarity is satisfied, which is equivalent to verifying Propositions~\ref{prop:karma-eq-stationarity-inequality}--\ref{prop:karma-eq-stationarity-equality}.
By construction of $\eta^\star_{i,j}(u_i) = \frac{\left(\bar r_i - \bar r^0_i\right)^*}{w_i} \, \eta^*_{i,j}(u_i)$, and using the Nash-balance condition~\eqref{eq:Nash-is-KE}, we have
\begin{multline*}
    \frac{\sum_{i'\in \N} w_{i'}}{w_i \sum_{j' \in \M} b^\star_{j'} \, c_{j'}} \left(\left(\bar r_i - \bar r^0_i\right)^\star - \sum\limits_{u'_i \in \U_i} \sum\limits_{j' \in \M} \eta^\star_{i,j'}(u'_i)\right) b^\star_j + \cfrac{\eta^\star_{i,j}(u_i)}{\sigma_i(u_i)} \\
    = \frac{\left(\bar r_i - \bar r^0_i\right)^*}{w_i} \left(\frac{\sum_{i'\in \N} w_{i'}}{\sum_{j' \in \M} \lambda^*_{j'} \, c_{j'}} \, (1 - C) \, \lambda^\star_j + \frac{\eta^*_{i,j}(u_i)}{\sigma_i} \right) \\
    = \frac{\left(\bar r_i - \bar r^0_i\right)^*}{w_i} \left(\lambda^*_j + \cfrac{\eta^*_{i,j}(u_i)}{\sigma_i(u_i)} \right).
\end{multline*}
Propositions~\ref{prop:karma-eq-stationarity-inequality}--\ref{prop:karma-eq-stationarity-equality} then follow from Propositions~\ref{prop:max-Nash-stationarity-inequality}--Propositions~\ref{prop:max-Nash-stationarity-equality}.
This concludes the proof.

\qed

\subsection{Proof of Proposition~\ref{prop:coupling}}
\label{app:proof-coupling}

By Proposition~\ref{prop:karma-eq-fair-share}, each user $i \in \N$ receives its fair share of each resource $j \in \{1,2\}$ in the separate economies, i.e., $\sum_{u_i \in \U_i} \sigma(u_i) \, \chi^{\star,j}_{i,j}(u_i) = \frac{w_i}{\sum_{i' \in \N}w_{i'}} \, c_j$.
This remains feasible for Problems~\eqref{eq:karma-user-problem} under the combined economy, irrespective of the equilibrium stationary clearing bids $\bm{b^\star}$:
\begin{multline*}
    \sum_{u_i \in \U_i} \sigma_i(u_i) \sum_{j \in \{1,2\}} \chi^{\star,j}_{i,j}(u_i) \, b^\star_j - \frac{w_i}{\sum_{i' \in \N} w_{i'}} \sum_{j \in \{1,2\}} b^\star_j \, c_j = \\ \sum_{j \in \{1,2\}} b^\star_j \left(\sum_{u_i \in \U_i} \sigma(u_i) \, \chi^{\star,j}_{i,j}(u_i) - \frac{w_i}{\sum_{i' \in \N}w_{i'}} \, c_j\right) = 0.
\end{multline*}
Therefore, the total reward under separate economies is a lower-bound for the total reward in the combined economy.

\qed